\documentclass[floats,aps,epsf,amsfonts]{revtex4}
\usepackage{ifpdf}
\usepackage[utf8]{inputenc}
\usepackage[UKenglish]{babel}
\usepackage{graphicx}
\usepackage[colorlinks]{hyperref}
\usepackage{amsmath, amsthm, amssymb}
\usepackage{multirow}
\usepackage{url}
\usepackage{color}

\definecolor{CiteColor}{rgb}{0,0.5,0}
\hypersetup{citecolor=CiteColor}
\definecolor{RefColor}{rgb}{0.55,0,0}
\hypersetup{linkcolor=RefColor}
\definecolor{darkgreen}{rgb}{0.2,0.7,0.2}

\newcommand{\pert}[1] {\bar{#1}}					
\newcommand{\diff}[2]  {\frac{d #1}{d #2}}
\newcommand{\sdiff}[2]  {\frac{d^2 #1}{d #2^2}}
\newcommand{\pdiff}[2]  {\frac{\partial #1}{\partial #2}}

\newcommand{\isco}{{\text{is}}}
\newcommand{\Feff}{\mathcal{F}_{\text{eff}}}
\newcommand{\cons}{{\text{cons}}}
\newcommand{\en}{\mathcal{E}}
\newcommand{\ang}{\mathcal{L}}
\newcommand{\pen}{\pert{\mathcal{E}}}
\newcommand{\pang}{\pert{\mathcal{L}}}
\newcommand{\pomegar}{\pert{\omega}_r}

\renewcommand{\c}{\hskip0.1cm,}
\newcommand{\p}{\hskip0.1cm.}
\renewcommand{\ell}{{\hat{l}}}

\mathchardef\mhyphen="2D

\def\prd{Phys. Rev. D}
\def\apj{Ap. J.}
\def\etal{\textit{et al.}}

\begin{document}

\title{Self force on a scalar charge in Kerr spacetime: eccentric equatorial orbits}
\author{Niels Warburton}
\author{Leor Barack}
\affiliation{School of Mathematics, University of Southampton, Southampton SO17 1BJ, United Kingdom}

\begin{abstract}

We present a numerical code for calculating the self force on a scalar charge moving in a bound (eccentric) geodesic in the equatorial plane of a Kerr black hole. We work in the frequency domain and make use of the method of extended homogeneous solutions [Phys.\ Rev.\ D {\bf 78}, 084021 (2008)], in conjunction with mode-sum regularization. Our work is part of a program to develop a computational architecture for fast and efficient self-force calculations, alternative to time-domain methods. We find that our frequency-domain method outperforms existing time-domain schemes for small eccentricities, and, remarkably, remains competitive up to eccentricities as high as $\sim 0.7$. As an application of our code we (i) compute the conservative scalar-field self-force correction to the innermost stable circular equatorial orbit, as a function of the Kerr spin parameter; and (ii) calculate the variation in the rest mass of the scalar particle along the orbit, caused by the component of the self force tangent to the four-velocity.

\end{abstract}	
\date{\today}
\maketitle

\section{Introduction}	

The last few years have seen a breakthrough in the program to compute the gravitational self-force (SF) on massive particles in orbit around black holes (see, e.g., \cite{Barack-Sago-eccentric} and references therein). Such systems are employed as models for strongly gravitating astrophysical binaries of extreme mass ratios (compact stellar objects captured by massive black holes in galactic nuclei), which are key sources of gravitational waves for planned detectors. The long-term ambition of the SF program is to model the phase evolution of the inspiral system, and the emitted gravitational waveforms, for generic inspiral orbits about rotating (Kerr type) black holes. However, to date, all computations of the gravitational SF have assumed a non-rotating (Schwarzschild type) central hole. The generalization to Kerr spacetime presents a major technical challenge, which the community must now come to address.

In a recent paper \cite{Warburton-Barack} (hereafter ``Paper I'') we reported a first computation of the SF for an orbit around a Kerr black hole. In this computation we resorted, as often in the SF program, to the simple framework of a scalar-charge toy model, and considered the scalar-field SF (SSF) acting on such a particle as it moves along a (fixed) circular equatorial geodesic of the Kerr geometry. The scalar-charge model provides a convenient environment for development, while already capturing much of the technical complexity of the full gravitational problem.

In Paper I we chose to take a frequency domain (FD) approach to the SF problem: The (scalar) field equations are decomposed into Fourier-harmonic modes on the Kerr background, and the resulting fully-separated ordinary differential equations (ODEs) are integrated numerically with suitable boundary conditions, in what is a routine procedure in black-hole perturbation studies. The Fourier-harmonic modes are then post-processed and used as input for the standard mode-sum regularization formula \cite{Barack-Ori} that yields the SSF. The FD approach has the obvious advantage that its numerical component only involves the solution of ODEs. As we have seen in Paper I, this makes it extremely efficient computationally in comparison with existing time-domain (TD) algorithms, which involve the numerical evolution of partial differential equations (PDEs). In the circular-orbit case, typical run-times of a current-day TD code (to compute the SF at a single radius) are of order hours, whereas an FD code arrives at the answer in seconds. The performance of the FD method is expected to deteriorate with increasing orbital eccentricity, as the Fourier spectrum of the wave equation broadens \cite{Barton-etal}, but, as we shall see in the current work, the method remains an efficient alternative to TD even for eccentricities as large as $\sim 0.7$.

Two main potential drawbacks of the FD approach, in view of the ultimate goals of the SF program, are the following. First, it is not obvious how an FD scheme might be implemented in an efficient way in a self-consistent evolution scheme that solves the field equations simultaneously with the equation of motion in order to track the self-forced evolution of the orbit. While TD algorithms can, in principle, feed the SF information back into the equation of motion ``in real time'', it is yet to be seen how FD schemes could accommodate a slowly evolving spectrum. One could possibly use FD data within the ``osculating geodesics'' approach to the orbital evolution (see \cite{Pound-Poisson:Osculating_orbits,Gair-etal}), which requires as input the value of the SF along each member of a sufficiently dense sequence of fixed geodesics. At any rate, FD codes can provide an extremely useful machinery for testing the results of TD algorithms. They can be used to produce a large amount of data quickly and accurately, and the core methods they rely on are entirely independent from those of the TD codes.

The second potential drawback becomes apparent when one moves on to consider the gravitational SF problem on the Kerr background. To the best of our knowledge, there is no known way to fully decompose the Lorenz-gauge perturbation equations in Kerr into Fourier-harmonic modes. It might be possible to proceed by using the tensorial spherical-harmonic decomposition as in the Schwarzschild problem  \cite{Barack-Lousto-2005,Barack-Sago-circular,Barack-Sago-eccentric,Akcay-Barack} and then properly account for the coupling between different harmonics that would occur in the Kerr case. An alternative would be to follow the radiation-gauge strategy advocated by Friedman {\it et al.}~\cite{Shah-etal, Keidl-etal}, which is naturally designed for an FD treatment.

In the current work we remain focused on the scalar field problem, and present an extension of the analysis of Paper I from circular (equatorial) orbits to eccentric (equatorial) orbits. This generalization is highly non-trivial in an FD treatment. The move from a single fundamental frequency (in the circular orbit case) to a bi-periodic motion brings with it several new technical complications (see below), which must be addressed. We develop here computational tools to deal with these complications; these tools should be transferable to the gravitational problem. This work represents a first implementation of the standard mode-sum regularization technique \cite{mode-sum-orig,Barack-review} for non-circular orbits in Kerr. As such, it contains, as a secondary result, a first numerical confirmation of the regularization parameter values for such orbits (these analytic parameters were derived long ago \cite{Barack-Ori} but were never used in an actual computation thus far).

Eccentric-orbit calculations in the FD, in both Schwarzschild and Kerr spacetimes, are hampered by the poor convergence of the Fourier sum near the particle singularity. The mode-sum formula requires as input the spherical-harmonic $l$-modes of the retarded field (and their derivatives), evaluated at the particle and summed over all $\omega$-frequencies. A na\"ive attempt to reconstruct the $l$-mode derivatives---which are generally discontinuous at the particle---as a sum of smooth Fourier harmonics leads to Gibbs-type oscillations, which severely impair the convergence of the Fourier sum in the vicinity of the particle (and, in fact, yields the wrong value at the particle itself). This problem has been identified and analyzed in Ref.\ \cite{Barack-Ori-Sago}, where a simple solution that entirely circumvents the problem was proposed. Using the proposed technique, dubbed the {\it method of extended homogeneous solutions} (EHS), the physical (retarded) field and its derivatives are reconstructed as an exponentially convergent sum over certain homogeneous solutions of the field equation. This technique has already been applied successfully in several numerical studies of perturbations from point particles \cite{Barack-Ori-Sago,Barack-Sago-eccentric,Hopper-Evans}, but here we implement it for the first time in a full computation of the SF. For that reason, we will take some time, in Sec. \ref{sec:efficiency}, to assess the computational performance of EHS.

A second challenge in extending the analysis of paper I to eccentric orbits has to do with a certain feature of the mode-sum scheme already mentioned above: the scheme requires as input the $l$-modes of the field (and its derivatives) in a \textit{spherical}-harmonic decomposition. Unfortunately, the field equations in Kerr spacetime do not separate into spherical harmonics, instead separating into \textit{spheroidal}-harmonic modes (in the FD). This means that the numerically-computed spheroidal modes must be post-processed in a procedure that projects them onto a basis of spherical harmonics. In paper I we demonstrated the manageability  of such a procedure in the circular-orbit case, where the spheroidicities encountered (proportional to the square of the modal frequencies) are relatively small, leading to a relatively weak coupling between the spheroidal and spherical modes. A potential concern is that the coupling might become less tractable with increasing eccentricity, since high harmonics of the orbital frequencies become more important, and these have higher spheroidicities leading to stronger coupling.  We show here, however, that even at moderately large eccentricities the coupling remains weak enough to remain manageable in practice.

Calculations of the SF for eccentric orbits give access to some interesting information about the post-geodesic dynamics in the binary system. In the gravitational problem, SF results have recently been used to quantify the precession effects of the conservative piece of the SF, and derive the resulting shift in the location and frequency of the innermost stable circular orbit (ISCO) \cite{Barack-Sago-ISCO-shift, Barack-Damour-Sago, Barack-Sago-precession}. In the SSF problem, the ISCO shift in Schwarzschild spacetime was derived by Diaz-Rivera \etal\ \cite{Diaz-Rivera}. Here we shall use our eccentric-orbit code to explore two effects of the SSF. First, we shall extend the work of Ref.\ \cite{Diaz-Rivera} to the Kerr case, and compute the conservative  shift in the location and frequency of the ISCO (for equatorial orbits) as a function of the Kerr spin parameter $a$. Second, we will compute the variation in the particle's rest mass as it moves along the eccentric orbit. This variation is caused by the component of the SSF tangent to the particle's four velocity. We will verify that this effect does not lead to a net change in the rest mass over a full orbital revolution, as expected for the configuration at hand.

The remainder of this paper is structured as follows. In section \ref{sec:setup} we review the relevant features of eccentric equatorial geodesics in Kerr geometry, and describe the formalism governing scalar-field perturbations. In section \ref{sec:mode-sum} we discuss the application of the mode-sum scheme to orbits in Kerr spacetime. In section \ref{sec:high-freq-problem} we examine the difficulties encountered with the na\"ive FD approach to SF calculations for eccentric orbits and review the resolution to these problems using the method of EHS. In section \ref{sec:numerics} we provide details of our numerical scheme and in section \ref{sec:results} we present a sample of our results and discuss the computational performance of the method. Sections \ref{sec:ISCO-shift} and \ref{sec:masschange} contain analyses of the conservative ISCO shift and of the rest-mass variation, respectively. Lastly in section \ref{sec:conclusion} we summarize our results and consider future work. Throughout this paper we use Boyer-Lindquist (BL) coordinates $(t,r,\theta,\varphi)$ with metric signature $(-+++)$ and geometrized units such that the gravitational constant $G$ and the speed of light $c$ are equal to unity.

\section{Preliminaries}\label{sec:setup}

\subsection{Orbital setup and equation of motion}\label{sec:orbital-setup}

Consider a pointlike particle of mass $\mu$ carrying a small quantity of scalar charge $q$ and moving on a bound orbit about a Kerr black hole of mass $M\gg\mu$ and spin $aM$. We assume that the correction to the background Kerr spacetime from the stress-energy of the particle's scalar field can be neglected. We denote the particle's worldline (in Boyer-Lindquist coordinates) by $x_p^\mu(\tau)$ and its four-velocity by $u^\alpha(\tau)=dx^\alpha(\tau)/d\tau$, where $\tau$ is the proper time. In this work we ignore the gravitational SF and seek to calculate only the SSF [of $O(q^2)$]. The orbital dynamics is described by the (self-)forced equation of motion \cite{Quinn}
\begin{eqnarray}\label{eq:motion}
	 u^\beta 	\nabla_\beta (\mu u^\alpha) = q \nabla^\alpha \Phi^R \equiv F^\alpha_{\text{self}} 		\c
\end{eqnarray}
where $\Phi^R$ is Detweiler--Whiting's smooth, regularized field \cite{Detweiler-Whiting}, and the covariant derivative is taken with respect to the background Kerr metric. In this work we calculate the SSF $F^{\alpha}_{\text{self}}$ ($ \propto q^2$) that would be felt by a particle moving on a fixed geodesic orbit, envisaging that this information could later be used to compute the true inspiralling orbit using a scheme similar to that described in \cite{Pound-Poisson:Osculating_orbits}. In this work we only consider the case of bound equatorial ($\theta_p=\pi/2$) orbits. 

As discussed in Paper I, once initial conditions are specified, a geodesic orbit about a Kerr black hole is uniquely parametrized by three constants of motion: its specific energy $\en\equiv-u_t$, angular momentum $\ang\equiv u_\varphi$ and Carter constant $\mathcal{Q}$. In the case of equatorial orbits the Carter constant vanishes and thus the pair $(\en,\ang)$ suffices to specify the orbit. It turns out to be convenient to use a different pair of parameters consisting of the semi-latus rectum $p$ and eccentricity $e$, representing strong-field generalizations of their Keplerian counterparts. For eccentric orbits we denote the BL radius at the point of closest approach (periastron) and the BL radius when the two bodies are furthest apart (apastron) by $r_{\text{min}}$ and $r_{\text{max}}$, respectively. Then $p$ and $e$ are defined by
\begin{eqnarray}
	p	\equiv	\frac{2r_\text{max}r_\text{min}}{r_\text{max} + r_\text{min}} \c \qquad e \equiv \frac{r_\text{max} - r_\text{min}}{r_\text{max} + r_\text{min}}		\p
\end{eqnarray}
The relation between $(p,e)$ and the specific energy and angular momentum is found to be \cite{Glampedakis-Kennefick}			
\begin{eqnarray}
	\en 		&=& 	\left [ 1 - \left( \frac{M}{p} \right) (1-e^2) \left\{1 - \frac{x^2}{p^2}(1-e^2) \right\} \right]^{1/2}				\c	\label{eq:energy}	\\
	\ang		&=&		x + a \en			\label{eq:ang-mom}		\c
\end{eqnarray}
where $x=x(a,p,e)$ is a rather complicated function given in appendix \ref{sec:equatorial-geodesics}.
All bound (stable) equatorial geodesics have $p^2>x^2 (1+e)(3-e)$; the ($a$ dependent) curve $p^2=x^2 (1+e)(3-e)$ defines a {\em separatrix} in the $e$--$p$ plane (see, e.g., Fig.\ 2 of \cite{Glampedakis-Kennefick}), with all orbits below the separatrix being unstable. For each given $a$, the intersection of the separatrix and the axis $e=0$ defines the ISCO.

Using the $p,e$ parametrization, the particle's orbital radius in BL coordinates can be expressed as
\begin{eqnarray}\label{eq:r_p}
	r = r_p (\chi) = \frac{p}{1+ e \cos \chi}	\c
\end{eqnarray}
where $\chi$ is a monotonically-increasing parameter along the particle's worldline. The azimuthal angle $\varphi_p$ and coordinate time $t_p$ can be computed as functions of $\chi$ along the particle's worldline using the expressions given in appendix \ref{sec:equatorial-geodesics}. We take $\varphi_p$ to be monotonically increasing for all orbits (or, equivalently, $\ang > 0$) and distinguish between prograde and retrograde orbits by the sign of $a$ ($a>0$ for prograde, $a<0$ for retrograde).
Without loss of generality we let $t_p(\chi=0)$ be the time of a periastron passage and denote the radial period (the $t$-time taken for the particle to progress from a periastron passage to a subsequent one) by $T_r \equiv t_p(\chi=2\pi) = 2 t_p(\chi=\pi) $. Similarly we denote the change in $\varphi_p$ over a time period $T_r$ by $\Delta\varphi_p \equiv \varphi_p(\chi=2\pi) = 2\varphi_p(\chi=\pi)$. Then the two frequencies associated with the radial and azimuthal motion are given by
\begin{eqnarray} \label{eq:fundamental-freqs}
	\Omega_r = \frac{2\pi}{T_r} \c \qquad \Omega_\varphi = \frac{\Delta\varphi_p}{T_r}	\p
\end{eqnarray}

Note that in Eq.\ \eqref{eq:motion} we have kept the mass term $\mu$ inside the derivative operator. Expanding the derivative one finds a term orthogonal to the particle's four-velocity, which is responsible for the self-acceleration, and a term tangential to the four-velocity, which in general gives rise to a dynamically varying mass. The orthogonal and tangential components of Eq.\ \eqref{eq:motion} are given respectively by
\begin{eqnarray}
\mu\diff{u^\alpha}{\tau} 	&=& (\delta^\alpha_\beta + u^\alpha u_\beta)F^\beta_{\text{self}} 	\equiv F^\alpha_{\perp(\text{self})}		\c \label{eq:orthogonal-force}	\\
\diff{\mu}{\tau} 			&=& - u^\alpha F_\alpha^{\text{self}} \p				\label{eq:mass-change}
\end{eqnarray}
Combining Eqs.\ \eqref{eq:motion} and \eqref{eq:mass-change} allows us to write the mass change explicitly as a function of $\tau$,
\begin{eqnarray}\label{eq:mass-change-explicit}
	\mu(\tau) = \mu_0 - q \Phi^R(\tau)	\c
\end{eqnarray}
where $\mu_0$ is a constant of integration (sometimes called the bare mass). In the case of circular, equatorial orbits discussed in Paper I there was no change in the particle's rest mass due to the stationarity of the setup. For eccentric orbits this is no longer the case. However, since $\Phi^R(\tau)$ comes back to itself after a radial period $T_r$, we expect no net mass change over that period.

\subsection{Field equation and multipole decomposition}

We assume that the field $\Phi$ associated with the particle's scalar charge $q$ obeys the minimally coupled Klein-Gordon equation
\begin{eqnarray}\label{eq:fieldeq}
\nabla^\alpha \nabla_\alpha \Phi = -4\pi T 			\c
\end{eqnarray}
where $T$ denotes the particle's scalar charge density. We model the latter as 
\begin{eqnarray}\label{T}
T = q \int \delta^4 ( x^\mu - x_p^\mu(\tau) ) [-g(x)^{-1/2}]  d\tau = \frac{q}{r_p^2 u^t} \delta(r-r_p)\delta(\varphi - \varphi_p)\delta(\theta - \pi/2)		\c
\end{eqnarray}
where $g= - \rho^4 \sin^2 \theta$ (with $\rho^2 \equiv r^2 + a^2 \cos^2 \theta$) is the determinant of the background Kerr metric, and in the second equality we have specialized to equatorial orbits with $\theta_p = \pi/2$. The $t$ component of the four-velocity, $u^t$, is calculated as $u^t = g^{t\varphi} \ang - g^{tt}\en$, where $g^{\alpha\beta}$ are the contravariant components of the Kerr metric tensor in BL coordinates, evaluated at the particle.

The scalar wave equation \eqref{eq:fieldeq} in Kerr geometry can be completely separated into spheroidal harmonic and frequency modes in the form \cite{Carter, Brill} 
\begin{eqnarray}\label{eq:field-decomp}
    \Phi = \int\sum_{\ell=0}^\infty\sum_{m=-\ell}^\ell R_{\ell m\omega}(r)S_{\ell m}(\theta;\sigma^2) e^{im\varphi} e^{-i\omega t}\, d\omega   \p
\end{eqnarray}
Here $S_{\ell m}(\theta;\sigma^2)$ are spheroidal Legendre functions [see Eq.\ (\ref{eq:angulareq}) below] with spheroidicity 
\begin{equation}
\sigma^2=-a^2\omega^2.
\end{equation} 
We reserve the term {\it spheroidal harmonic} for the product $S_{\ell m}(\theta;\sigma^2)e^{im\varphi}$. Notice that we label spheroidal-harmonic modes by $\ell m$ as we will later introduce {\em spherical}-harmonics modes which we will label by $lm$. The spheroidal harmonics are orthonormal with normalization given by
\begin{eqnarray}\label{eq:spheroidal-harmonic-normalization}
    \oint S_{\ell m}(\theta;\sigma^2)e^{im\varphi}S_{\ell' m'}(\theta;\sigma^2)e^{-im'\varphi} \, d\Omega = \delta_{\ell\ell'} \delta_{mm'} \c 
\end{eqnarray}
where $d\Omega=\sin\theta d\theta d\varphi$, and $\delta_{n_1n_2}$ is the standard Kronecker delta. 

The source $T$  in Eq.\ (\ref{T}) has a discrete spectrum given by 
\begin{eqnarray}\label{eq:omega}
	\omega = \omega_{mn} = m \Omega_\varphi + n \Omega_r	\c
\end{eqnarray}
for integer $m$ and $n$. It therefore admits a discrete Fourier decomposition, of the form
\begin{eqnarray}\label{eq:source-decomp}
     \rho^2 T = \sum_{\ell=0}^\infty\sum_{m=-\ell}^\ell \sum_{n=-\infty}^{\infty} {T}_{\ell mn}(r)S_{\ell mn}(\theta) e^{im\varphi} e^{-i\omega_{mn} t}  \c
\end{eqnarray}
where $S_{\ell mn}(\theta)\equiv S_{\ell m}(\theta;-a^2\omega_{mn}^2)$, and the factor $\rho^2$ is introduced for later convenience. Using the orthonormality property  \eqref{eq:spheroidal-harmonic-normalization} and taking the inverse Fourier transform of Eq.\ \eqref{eq:source-decomp} we find
\begin{eqnarray}
	 {T}_{\ell mn}(r) = q S_{\ell mn}(\pi/2) T_r^{-1} \int_0^{T_r} (u^t)^{-1} \delta[r-r_p(t)]e^{i[\omega_{mn} t-m\varphi_p(t)]} \, dt	\c
\end{eqnarray}
where $\varphi_p(t) \equiv \varphi_p(\chi(t))$, and $\chi(t)$ is obtained formally by inverting $t_p(\chi)$  (this inverse exists since $t$ is a monotonically increasing function of $\chi$). Noting that ${T}_{\ell mn}(r)$ only has support on $r_\text{min} \le r \le r_\text{max}$ and changing the integration variable from $t$ to $r_p$ [taking $t_p(\chi=0)=\varphi_p(\chi=0)=0$ without loss of generality ], we finally obtain
\begin{eqnarray}\label{eq:source-term}
	 {T}_{\ell mn}(r) = \frac{2q S_{\ell mn}(\pi/2)}{T_r |u^r(r)|} \cos[\omega_{mn} t_p(r)- m \varphi_p(r)] \times \Theta(r-r_\text{min}) \times \Theta (r_\text{max} - r) \c
\end{eqnarray}
where $\Theta$ is the standard Heaviside step function, and $t_p(r)$ and $\varphi_p(r)$ are obtained by formally inverting $r_p(\chi)$ in the range $0 \le \chi \le \pi$. 

The decompositions (\ref{eq:source-decomp}) and (the discrete version of) \eqref{eq:field-decomp} separate the $r$ and $\theta$ dependence of the field equation (\ref{eq:fieldeq}), with resulting ${\ell mn}$-mode equations given by 
\begin{widetext}
\begin{equation}
    \Delta\pdiff{}{r}\left(\Delta\pdiff{R_{\ell m n}}{r}\right) + \left[a^2m^2-4Mrma\omega_{mn} + (r^2+a^2)^2\omega_{mn}^2 -a^2\omega_{mn}^2\Delta - \lambda_{\ell m}\Delta)\right]R_{\ell m n} = -4\pi\Delta  {T}_{\ell mn}(r)    \label{eq:radialeq} \c
\end{equation}
\begin{equation}
    \frac{1}{\sin\theta}\pdiff{}{\theta}\left(\sin\theta\pdiff{S_{\ell mn}}{\theta}\right) + \left(\lambda_{\ell m} + a^2\omega_{mn}^2\cos^2\theta - \frac{m^2}{\sin^2\theta}\right)S_{\ell mn} = 0 \c \label{eq:angulareq}
\end{equation}
\end{widetext}
where $\Delta\equiv r^2-2Mr+a^2$ and $R_{\ell m n}\equiv R_{\ell m \omega_{mn}}$. The angular equation (\ref{eq:angulareq}) takes the form of the spheroidal Legendre equation with spheroidicity $\sigma^2=-a^2\omega_{mn}^2$. Its eigenfunctions are the (frequency dependent) spheroidal Legendre functions $S_{\ell mn}(\theta;\sigma^2)$, with corresponding eigenvalues $\lambda_{\ell m}$.  In general there are no closed-form expressions for $S_{\ell mn}$ or $\lambda_{\ell m}$ but they can be calculated using the spherical-harmonic decomposition method described in appendix A of Paper I.  For $\sigma^2=0$ the spheroidal harmonics $S_{\ell mn}e^{i m \varphi}$ become the standard spherical harmonics $Y_{\hat lm}$, and their eigenvalues reduce to $\lambda_{\ell m} =\ell(\ell+1)$.

To simplify the construction of boundary conditions (below), and assist our numerical scheme, it is convenient to transform to a new radial variable,
\begin{eqnarray}
    \psi_{\ell m n}(r)\equiv rR_{\ell m n}(r)       \c          \label{eq:R-psi}
\end{eqnarray}
and introduce the tortoise radial coordinate $r_*$ defined through
\begin{equation}\label{eq:rs}
    \diff{r_*}{r} = \frac{r^2}{\Delta} \p
\end{equation}
With the above definition the tortoise coordinate is given explicitly in terms of $r$ as
\begin{eqnarray}\label{eq:tortoise-explicit}
    r_* = r + M \ln(\Delta/M^2) + \frac{(2M^2 - a^2)}{2(M^2-a^2)^{1/2}}\ln\left(\frac{r-r_+}{r-r_-}\right)  \c
\end{eqnarray}
where we have specified the constant of integration, and where $r_\pm = M \pm \sqrt{M^2-a^2}$ are the outer and inner roots respectively of the equation $\Delta=0$. There are several standard alternative choices that can be made for the tortoise coordinate. Our choice is motivated by the observation of Bardeen \etal\ \cite{Bardeen} that it leads to a simpler radial potential than other common choices \cite{Brill}. This, in particular, simplifies the construction of the numerical boundary conditions for the resulting radial equation.
In terms of $\psi_{\ell m n}(r)$ and $r_*$, the radial equation (\ref{eq:radialeq}) takes the simpler form
\begin{widetext}
\begin{eqnarray}\label{eq:rsradialeqn}
   \sdiff{\psi_{\ell m n}}{r_*} + V_{\ell m n}(r)\psi_{\ell m n} =  -\frac{4 \pi\Delta}{r^3} {T}_{\ell mn}(r)  \equiv Z_{\ell mn}(r) \c
\end{eqnarray}
where ${T}_{\ell m \omega}$ is given in Eq.\ \eqref{eq:source-term} above, and $V_{\ell m n}$ is an effective ($\omega$-dependent) radial potential given by
\begin{eqnarray}
    V_{\ell m n}(r) = \left[\frac{(r^2+a^2)\omega_{mn} - am}{r^2}\right]^2 - \frac{\Delta}{r^4}\left[\lambda_{\ell m} - 2am\omega_{mn} + a^2\omega_{mn}^2 + \frac{2(Mr-a^2)}{r^2}\right]\p
\end{eqnarray}
There is no known closed-form analytic solution to the radial equation \eqref{eq:rsradialeqn} for general $\ell m n$ and so it has to be solved numerically. Our technique for doing this is detailed in section \ref{sec:numerics}.
\end{widetext}

\subsection{Boundary conditions}\label{sec:asymptotic-bcs}

The solution of the radial equation \eqref{eq:rsradialeqn} is uniquely determined outside the black hole once suitable boundary conditions are specified on the horizon $(r_* \to -\infty)$ and at spatial infinity $(r_* \to \infty)$. In order to select the retarded solution to the field equation \eqref{eq:fieldeq} we require that radiation be `outgoing' at infinity and `ingoing' at the horizon. Making this statement precise is straightforward at spatial infinity but there is some subtlety at the event horizon following from the dragging of inertial frames in Kerr spacetime. We go though the calculation for both boundaries in section II.C. of Paper I and as the boundary conditions have no dependence on the type of orbit being examined all that is said there carries through here, with the only change being that $\omega$ is now bi-harmonic in $m$ and $n$. Here it will suffice to simply state the results presented in Paper I. The asymptotic boundary conditions for the radial equation at spacial infinity and the event horizon are given, respectively, by
\begin{eqnarray}
	\psi_{lmn} (r_* \to \infty ) &\sim& e^{+i \omega_{mn} r_*}		\c		\label{eq:asym-bcs-inf}\\
	\psi_{lmn} (r_* \to -\infty ) &\sim& e^{-i \gamma_{mn} r_*}	\c		\label{eq:asym-bcs-horiz}
\end{eqnarray}
where
\begin{eqnarray}
\gamma_{mn} =  \frac{2Mr_+ \omega_{mn} - am}{r_+^2}		\p
\end{eqnarray}
These boundary conditions specify a ratio between the value of the radial function and its derivative. Later on, in section \ref{sec:numerical-bcs}, we will set an (arbitrary) amplitude scaling for the homogeneous solutions. The unique inhomogeneous solutions will then be constructed via the method of EHS, as described in section \ref{sec:EHS}

\section{Self-force via mode-sum regularization}\label{sec:mode-sum}

In the mode-sum scheme the full field is decomposed into spherical-harmonic modes, even in Kerr spacetime. We define the \textit{full force} at an arbitrary field point $x\equiv x^\mu$ in the neighbourhood of our scalar charge $q$ as the field
\begin{eqnarray}
	F^{\text{full}}_\alpha (x) \equiv q \nabla_\alpha \Phi(x)	= \sum_l F_\alpha^\text{(full)l} (x)	\c
\end{eqnarray}
where $\Phi(x)$ is the physical (retarded) scalar field, and $F_{\alpha}^{\text{(full)}l}$ denotes the total contribution to $q\nabla_\alpha \Phi$ from its spherical harmonic $l$-mode (summed over $m$). Each $F_{\alpha}^{\text{(full)}l}$ is finite at the particle’s location, although in general the sided limits $r \rightarrow r_p^\pm$ yield two different values, denoted $F_{\alpha\pm}^{\text{(full)}l}$ respectively. The mode-sum formula for the SSF acting on the particle reads \cite{Barack-Ori}
\begin{eqnarray}\label{eq:ret-mode-sum}
	F^{\text{self}}_\alpha = \sum_{l=0}^\infty \left( F^{l\text{(full)}}_{\alpha \pm} - A_{\alpha\pm} (l+1/2) - B_\alpha \right)	\c
\end{eqnarray}
were the coefficients $A_{\alpha\pm}$ and $B_\alpha$ are $l$-independent ``regularization parameters'', the values of which are known analytically for generic orbits about a Kerr black hole \cite{Barack-Ori} (see Ref.\ \cite{Barack-review} for a full derivation of the regularization parameters in the Kerr case). We note the difference $F^{l\text{(full)}}_{\alpha \pm}(x_p) - A_{\alpha\pm} (l+1/2)$ no longer exhibits the $\pm$ ambiguity.

In Kerr, as we have seen, the field naturally decomposes into {\em spheroidal} harmonic modes. The mode-sum scheme, on the other hand, requires {\em spherical} harmonic modes as input even in Kerr spacetime. Hence in order to regularize using the standard mode-sum approach we first need to project the spheroidal-harmonic contributions onto a basis of spherical harmonics. We do this by expanding
\begin{eqnarray}\label{eq:spheroidal-decomp}
	S_{\ell m}(\theta; \sigma^2) e^{im\varphi} = \sum_{l=0}^{\infty} b^\ell_{lm}(\sigma^2) Y_{lm}(\theta,\varphi)		\c
\end{eqnarray}
where the $\sigma$-dependent coefficients $b^\ell_{lm}$ are determined from a recursion relation found by substituting the series expansion into the angular differential equation \eqref{eq:angulareq}---a procedure which follows \cite{Hughes} and is described in detail in Appendix A of Paper I. 
Given the coupling coefficients $b^\ell_{lm}$, we can write the spherical-harmonic (``$l$-mode'') contribution to the full force as 
\begin{eqnarray}\label{eq:field-spherical-mode}
    F_\alpha^{\text{(full)}l}(x) = q \nabla_\alpha \left[\sum_{m=-l}^l  \phi_{l m}(t,r) Y_{l m}(\theta,\varphi) /r  \right] \c
\end{eqnarray}
where 
\begin{eqnarray}\label{eq:phi_lm}
	\phi_{lm}(t,r) = \sum_{n=0}^\infty \sum_{\ell=|m|}^\infty b_{lm}^\ell \psi_{\ell m}(r) e^{-i \omega_{mn} t}		\p
\end{eqnarray}
In the last equation the convergence of the sum over $n$ is impaired near the particle due to the Gibbs phenomenon; we will discuss this problem and its resolution  in section \ref{sec:high-freq-problem} below.

Formally, when constructing $\phi_{lm}$ via Eq.\ \eqref{eq:phi_lm} one has to sum over an infinite number of spheroidal $\ell$ modes. In practice this is not necessary as the coupling between the spheroidal and spherical harmonic modes is relatively narrow-band for the spheroidicities encountered in our calculation. In Paper I we numerically demonstrated that the contribution from a given spheroidal $\ell m$ mode to the spherical harmonic $lm$ modes of the field is strongly peaked around $l=\ell$ with exponential wings. The coupling strengthens as the magnitude of the spheroidicity ($=a^2\omega^2$) increases. In the case of circular orbits we found that for the largest spheroidicity encountered in our work (for a black hole spin of $a=0.998M$ and orbital radius of $r_0=2M$) the coupling was still weak enough to be perfectly manageable in practice. For example, computing the first 50 spherical-harmonic mode contributions required only $\sim$56 spheroidal-harmonic modes (i.e. even in this extreme case the bandwidth of the coupling is still only $\pm$6 modes). In the case of eccentric equatorial orbits the situation is slightly worse as the spectrum is now bi-periodic ($\omega = m\Omega_\varphi + n \Omega_r$), and the relevant values of $\omega$, and consequently $\sigma^2$, can be much larger than in the circular orbit case. Nonetheless we find that the coupling is still weak enough to allow for reasonably efficient SSF computations (see section \ref{sec:efficiency} for details of the computational performance of the method). As an example, for an orbit with parameters $(a,p,e) = (0.9M,10M,0.5)$ (giving the highest spheroidicity considered in this work) we find, for $l=14$, an effective bandwidth of $\sim$12 modes. 

\subsection{Conservative and dissipative pieces of the SF}\label{sec:cons-diss}

The SF can be split into conservative and dissipative components \cite{Hinderer-Flanagan, Barack-review}. This split is useful for analysing the different physical effects of the SF. Splitting the SF into its conservative and dissipative components is also practically beneficial as the two pieces admit $l$-mode sums with different convergence properties, which are better dealt with separately. Let us write 
\begin{eqnarray}\label{eq:construct-SSF-from-diss-cons}
	F_\alpha^\text{self} \equiv F_\alpha^{\text{ret}} = F^{\text{cons}}_\alpha + F^{\text{diss}}_\alpha		\c
\end{eqnarray}
where 
\begin{eqnarray}\label{eq:cons-diss-ret-adv}
	F_\alpha^{\text{cons}} \equiv \frac{1}{2}(F_\alpha^{\text{ret}} + F_\alpha^{\text{adv}}), \qquad F_\alpha^{\text{diss}} \equiv \frac{1}{2}(F_\alpha^{\text{ret}} - F_\alpha^{\text{adv}})		\c
\end{eqnarray}
in which the \textit{advanced} SF is obtained from Eq.\ \eqref{eq:ret-mode-sum} by replacing $F^{\text{(full)}l}_\alpha(\Phi_\text{ret})$ with $F^{\text{(full)}l}_\alpha(\Phi_\text{adv})$, i.e.
\begin{eqnarray}\label{eq:adv-mode-sum}
	F_\alpha^\text{adv} = \sum_{l=0}^\infty [ F_{\alpha l\pm}^{\text{full}}(\Phi_{\text{adv}}) - A_{\alpha\pm} (l+1/2) - B_\alpha ]		\c
\end{eqnarray}
where $\Phi_\text{adv}$ is the advanced solution to Eq.\ \eqref{eq:fieldeq}. 	

Substituting into formulae \eqref{eq:cons-diss-ret-adv} from Eqs.\ \eqref{eq:ret-mode-sum} and \eqref{eq:adv-mode-sum}, we obtain mode-sum regularization formulae for the conservative and dissipative components \cite{Barack-review}:
\begin{eqnarray}
F_\alpha^{\text{cons}} &=& \sum_{l=0}^\infty \left[ F_{\alpha l \pm}^{\text{full(cons)}} - A_\alpha^\pm (l+1/2) - B_\alpha \right]	\c	\label{eq:cons-regularization}	\\
F_\alpha^{\text{diss}} &=& \sum_{l=0}^\infty F_{\alpha l \pm}^{\text{full(diss)}}													\c	\label{eq:diss-regularization}
\end{eqnarray}
where
\begin{eqnarray}
	F_{\alpha l\pm}^{\text{full(cons)}} \equiv \frac{1}{2} [ F_{\alpha l \pm}^{\text{full}}(\Phi_\text{ret}) + F_{\alpha l\pm}^\text{full}(\Phi_\text{adv})], \qquad F_{\alpha l \pm}^\text{full(diss)} \equiv \frac{1}{2} [ F_{\alpha l\pm}^\text{full}(\Phi_\text{ret})-F_{\alpha l \pm}^\text{full}(\Phi_\text{adv})]		\p
\end{eqnarray}
Note that the dissipative piece of the SF does not require regularization, whilst the conservative piece does. It turns out that the $l$-mode sum for the dissipative piece converges exponentially, while that for the conservative piece converges only as $\propto l^{-1}$ (the $l$-mode terms fall off as $\propto 1/l^2$ at large $l$) \cite{Barack-review}.

The advanced field $\Phi_{\text{adv}}$ can be constructed by reversing the boundary conditions (\ref{eq:asym-bcs-inf}) and (\ref{eq:asym-bcs-horiz}) so that radiation is ingoing at infinity and outgoing at the horizon. This, however, doubles the computational cost. Fortunately there is a simple way around this, which takes advantage of the particular symmetries of geodesics in Kerr (noted in a different context by Mino in \cite{Mino}): As explained in \cite{Barack-review}, it is possible to re-express Eq.\ \eqref{eq:cons-diss-ret-adv} in terms of the retarded force $F_\alpha^{\text{ret}}$ alone, by considering the value of $F_\alpha^{\text{ret}}$ at two ``opposite'' points along the equatorial orbit, i.e., ones with the same value of $r_p$ but opposite radial velocities. It is convenient to take $\tau=0$ to correspond to a periapsis passage, and express the SSF as a function of $\tau$ along the orbit. Then one can show  
\cite{Barack-review}
\begin{eqnarray} \label{eq:diss-cons}
F_\alpha^\text{cons}(\tau) = \frac{1}{2} [ F_\alpha^\text{ret}(\tau) + \epsilon_{(\alpha)} F_\alpha^\text{ret}(-\tau)], 	\qquad		F_\alpha^\text{diss}(\tau) = \frac{1}{2} [ F_\alpha^\text{ret} (\tau) - \epsilon_{(\alpha)} F_\alpha^\text{ret}(-\tau)] \c
\end{eqnarray}
where $\epsilon_{(\alpha)}=(-1,1,1,-1)$ in Boyer-Lindquist coordinates. Analogous relations apply for $F_{\alpha l\pm}^{\text{full(cons)}}$ and $F_{\alpha l\pm}^{\text{full(diss)}}$, and we shall use them in our analysis to construct the necessary input for the mode sums 
(\ref{eq:cons-regularization}) and (\ref{eq:diss-regularization}).

\section{Review of the high-frequency problem and its resolution}\label{sec:high-freq-problem}

This problem was first discussed in Ref.\ \cite{Barack-Ori-Sago}. We review it here for completeness, and where appropriate generalize the discussion to Kerr spacetime. 

Let $\psi_{\ell mn}^+(r)$ and $\psi_{\ell mn}^-(r)$ be two homogeneous solutions to the radial equation \eqref{eq:rsradialeqn}, satisfying the boundary conditions presented in Eqs.\ \eqref{eq:asym-bcs-inf} and \eqref{eq:asym-bcs-horiz}, respectively. These two homogeneous solutions form a basis that can be used to construct the physical inhomogeneous solution using the standard variation of parameters method:
\begin{eqnarray}\label{eq:Wronskian-formula}
\psi_{\ell mn}(r) &=& \psi^+_{\ell mn}(r) \int^r_{r_{\text{min}}} \frac{\psi^-_{\ell mn}(r') Z_{\ell mn}(r')r'^2}{\Delta(r')W } \, dr' + \psi^-_{\ell mn}(r) \int^{r_{\text{max}}}_r \frac{\psi^+_{\ell mn}(r')Z_{\ell mn}(r')r'^2}{\Delta(r')W } \, dr'		\c	\nonumber \\
				&\equiv& \psi^{\text{inh}}_{\ell mn}(r)								\p
\end{eqnarray}
Here we have changed the integration variable from $r_*$ to $r$ using Eq.\ \eqref{eq:rs}, and
\begin{eqnarray}
W \equiv \psi^-_{\ell mn} (d\psi^+_{\ell mn}/dr_*) - \psi^+_{\ell mn} (d\psi^-_{\ell mn}/dr_*) = \text{const} 
\end{eqnarray}
is the Wronskian. In the regions $r \leq r_{\text{min}}$ and $r \geq r_{\text{max}}$ this formula reduces to the homogeneous solutions
\begin{eqnarray}\label{eq:homogen-sols}
\psi_{lmn}(r) = \begin{cases}
					C^-_{\ell mn} \psi^-_{\ell mn} \equiv \tilde{\psi}^-_{\ell mn}(r), \qquad r \leq r_{\text{min}}	\c	\\
					C^+_{\ell mn} \psi^+_{\ell mn} \equiv \tilde{\psi}^+_{\ell mn}(r), \qquad r \geq r_{\text{max}}	\c 	\\
			\end{cases}
\end{eqnarray}
where the coefficients $C^-_{\ell mn}$ and $C^+_{\ell mn}$ are given by
\begin{eqnarray}\label{eq:scaling-coeffs}
	C^\pm_{\ell mn} = W^{-1} \int^{r_{\text{max}}}_{r_{\text{min}}} \frac{ \psi^\mp_{\ell mn}(r) Z_{\ell mn} (r) r^2}{\Delta(r)} \, dr 	\p
\end{eqnarray}
The source term $Z_{\ell mn}(r)$, defined in Eq.\ \eqref{eq:rsradialeqn}, has singularities at $r=r_\text{max},r_\text{min}$ and to avoid them it is convenient to change the integration variable from $r$ to $t_p(r)$. Taking $t_p(r_\text{min})=0$, the scaling coefficients $C^\pm_{\ell mn}$ are then given by
\begin{eqnarray}\label{eq:scaling-coefficients}
	C^\pm_{\ell mn} = -\frac{8 \pi q S_{\ell mn} (\pi/2)}{T_r W} \int_0^{T_r/2} \frac{\psi^\mp_{\ell m \omega}(r_p(t)) \cos(\omega_{mn} t - m \varphi_p(t))}{r_p(t) u^t(r_p(t))} \, dt 		\c
\end{eqnarray}
where now the integrand is free from singularities.

We now need to construct the time-domain field $\phi_{lm}(t,r)$ and its $t$ and $r$ derivatives. Recalling Eq.\ \eqref{eq:phi_lm}, this field is constructed via
\begin{eqnarray}
	\phi_{lm}(t,r) 			&=& \sum_n \phi_{lmn}(t,r)	\label{eq:phi-tr}	\c
\end{eqnarray} 
with
\begin{eqnarray}
	\phi_{lmn}(t,r)			&=&	\sum_{\ell=0}^\infty b_{lm}^\ell \psi^{\text{inh}}_{\ell mn}(r) e^{-i\omega_{mn} t}		\c
\end{eqnarray}
where the $b_{lm}^\ell$'s are the spheroidal-spherical harmonic coupling coefficients from Eq.\ \eqref{eq:spheroidal-decomp}. For any given $r_0$ between $r_\text{min}$ and $r_\text{max}$, the field $\phi_{lm}(t,r_0) $ is {\em not} a smooth function of time (its derivative suffers a jump discontinuity when the particle crosses $r_0$). Standard Fourier theory tells us that the sum over $n$ in Eq.\ \eqref{eq:phi-tr} will suffer Gibbs-type oscillations near the particle, and will converge very slowly as a result. Even more troubling, the derivatives of the field, $\phi_{lm,\alpha}$, which are needed as input for the mode-sum scheme, may fail to converge to the correct value at the discontinuity. One might attempt to extract the correct value of the derivatives through extrapolation, but such a procedure would be computationally inefficient because of the poor convergence of the $n$-mode sum in the vicinity of the particle.

It should be clarified that the above problem does not occur in the case of circular orbits, even if they are inclined, as for these orbits the field at any fixed radius is a smooth function of time.

\subsection{Method of extended homogeneous solutions}\label{sec:EHS}

The above problem might have made SF calculations in the frequency domain rather unattractive. Fortunately, Ref.\ \cite{Barack-Ori-Sago} recently proposed a simple technique for overcoming this problem, dubbed the method of \textit{extended homogeneous solutions} (EHS). Ref.\ \cite{Barack-Ori-Sago} outlined the details of the method and provided a numerical example using EHS to calculate the monopole contribution to the scalar field for a particle in an eccentric orbit about a Schwarzschild black hole. Here we will overview the method and provide the formula required to extend their calculation to eccentric equatorial orbits about a Kerr black hole.

The first step in the method (as its name suggests) is to consider an extension of the homogeneous solutions $\psi_{\ell m \omega}^\pm$ to the entire domain, defined through
\begin{eqnarray}\label{eq:EHS-omega-mode}
	\tilde{\psi}^\pm_{\ell mn}(r) \equiv C^\pm_{\ell mn} \psi^\pm_{\ell mn}(r), \qquad r > 2M	\c
\end{eqnarray}
where the coefficients $C^+_{\ell mn}$ and $C^-_{\ell mn}$ are those given in Eq.\ \eqref{eq:scaling-coeffs}. Note that $\tilde{\psi}^+_{\ell mn}$ and $\tilde{\psi}^-_{\ell mn}$ coincide with $\psi^{\text{inh}}_{\ell mn}$ in the respective domains $r > r_{\rm max}$ and $r < r_{\rm min}$, but neither is a solution of the inhomogeneous $n$-mode equation (\ref{eq:rsradialeqn}) in the sourced domain $r_{\rm min}\leq r\leq r_{\rm max}$.
One then defines the two spherical-harmonic time-domain extended homogeneous solutions $\tilde{\phi}^+_{lm}$ and $\tilde{\phi}^-_{lm}$ by
\begin{eqnarray}\label{eq:EHS}
	\tilde{\phi}_{lm}^\pm(t,r) \equiv \sum_n \tilde{\phi}_{lmn}^\pm(t,r)	\c
\end{eqnarray}
where
\begin{eqnarray}\label{eq:phi_lmn}
	\tilde{\phi}_{lmn}^\pm(t,r) = \sum_{\ell = 0}^\infty b_{lm}^\ell \tilde{\psi}_{\ell m n}^\pm(r)e^{-i\omega_{mn}t}		\p
\end{eqnarray}
Ref.\ \cite{Barack-Ori-Sago} showed that the $n$-mode sum in Eq.\ \eqref{eq:EHS} converges exponentially-fast in $|n|$, and that it does so uniformly in $t$ and $r$ throughout $r > 2M$ (note that $\tilde{\phi}_{lm}^\pm$ are smooth functions of $t$ for any $r$). Furthermore, Ref.\ \cite{Barack-Ori-Sago} argued that the extended homogeneous solutions can be used to construct the true time-domain function $\phi_{lm}(t,r)$ on either side of the particle's trajectory:
\begin{eqnarray}\label{eq:actual-time-domain-function}
\phi_{lm}(t,r) = \begin{cases}
				\tilde{\phi}^+_{lm}(t,r), \qquad r \geq r_p(t)  \c			\\
				\tilde{\phi}^-_{lm}(t,r), \qquad r \leq r_p(t)  \p
				\end{cases}
\end{eqnarray}
The SSF is then obtained via Eqs.\ \eqref{eq:field-spherical-mode} and \eqref{eq:ret-mode-sum} as usual.

The advantages of this method are clear. First, the problem of Gibbs ringing encountered when constructing the derivative of the field is no longer present; the derivative of the field converges with $n$ to the correct value everywhere, including at the location of the particle. Second, the convergence of both the field and its derivatives is exponential, making an EHS calculation extremely computationally efficient. Third, the standard Wronskian-based method requires the evaluation of two integrals [see Eq.\ \eqref{eq:Wronskian-formula}] {\em for each orbital point separately} (and for each $\ell m n$). In EHS, on the other hand, the two integrals in Eq.\ \eqref{eq:scaling-coeffs} suffice to determine the $\ell m n$-mode along the entire orbit.

Since the introduction of the method of EHS it has been applied to a range of problems, including the calculation of the monopole and dipole contributions to the gravitational SF in the Lorenz gauge \cite{Barack-Sago-eccentric}, and the calculation of the full gravitational perturbation in the Regge-Wheeler gauge  \cite{Hopper-Evans}. Our work represents a first implementation of EHS for a full calculation of the SF.

\section{Numerical implementation}\label{sec:numerics}

The radial equation \eqref{eq:rsradialeqn} has no known closed-form analytic solution for general $\ell m n$ and as such it needs to be solved numerically. For a given point in the $(a,p,e)$ parameter space, a calculation of the SSF will entail solving for many $\ell mn$ modes (see below for details). To assist the calculation we note that $\psi^\pm_{\ell m n}$ remains unchanged under the combined operations of $(m,n) \rightarrow (-m,-n)$ and complex conjugation. Consequently, we need only numerically solve for the $m\geq 0$ modes---although we still have to solve for both positive and negative $n$ modes (except for $m=0$). The computational burden is further reduced by noting that for $l+m=\text{odd}$ the spheroidal harmonics vanish at $\theta=\pi/2$ and as a result the corresponding $\psi^\pm_{\ell mn}$, as well as their $t,r,\varphi$ derivatives, automatically vanish at the particle (the $\theta$ derivative is also zero there, from symmetry).

\subsection{Numerical boundary conditions for the homogeneous fields}\label{sec:numerical-bcs}

The asymptotic boundary conditions for the radial equation are presented in Eqs.\ \eqref{eq:asym-bcs-inf} and \eqref{eq:asym-bcs-horiz}. However, in order to solve the radial equation numerically, boundary conditions need to be specified at finite radii. We denote the $r_*$ radius of the inner boundary by $r_{*\text{in}} \ll -M$ and that of the outer boundary by $r_{*\text{out}} \gg M$ (we discuss how these radii are chosen in practice in the algorithm section below). In order to determine the boundary conditions at $r_{*\text{in}}$ and $r_{*\text{out}}$ we use the ans\"atze
\begin{eqnarray}
	\psi^+_{\ell m \omega_{mn}} (r_\text{out}) &=& e^{+\omega_{mn} r_{*\text{out}}} \sum_{k=0}^{k_\text{out}} c_k^+ r^{-k}_\text{out}	\label{eq:inf-bc}		\c	\\
	\psi^-_{\ell m \omega_{mn}} (r_\text{in})	 &=& e^{-\gamma_{mn} r_{*\text{in}}}  \sum_{k=0}^{k_\text{in}}  c_k^{-}  (r_\text{in} - r_+)^k	\label{eq:horiz-bc}		\c
\end{eqnarray}
where $r_\text{in} = r(r_{*\text{in}})$, $r_\text{out} = r(r_{*\text{out}})$, and the truncation parameters $k_\text{in,out}$ are chosen such that the boundary conditions reach a prescribed accuracy (see algorithm section below). Substituting the above series into the radial equation \eqref{eq:rsradialeqn} gives recursion relations for the $c_{k>0}^{\pm}$ coefficients in terms of $c_0^{\pm}$, respectively. In practice we take $c_0^\pm = 1$. The explicit form of the recursion relations is rather unwieldy and can be found in appendix C of Paper I. 

In the case of the static $m=n=0$ modes there exist simple analytic solutions for $\psi^\pm_{\ell mn}$, determined by regularity at the event horizon and at spatial infinity; they read
\begin{eqnarray}\label{eq:m-n-equals-zero}
	\psi^-_{\ell 00} = \alpha_- r P_\ell(z)\c \qquad \psi^+_{\ell 00} = \alpha_+ r Q_\ell(z) \c
\end{eqnarray}
where $\alpha_-,\alpha_+$ are arbitrary constants, $P_\ell$ and $Q_\ell$ are the Legendre function of the first and second kind respectively, and
\begin{eqnarray}
	z \equiv \sqrt{ \frac{M^2+a^2}{M^4-a^4} }(r-M)	 		\p
\end{eqnarray}

\subsection{Algorithm}\label{sec:algorithm}

\begin{figure}
	\begin{center}
	\includegraphics[width=80mm]{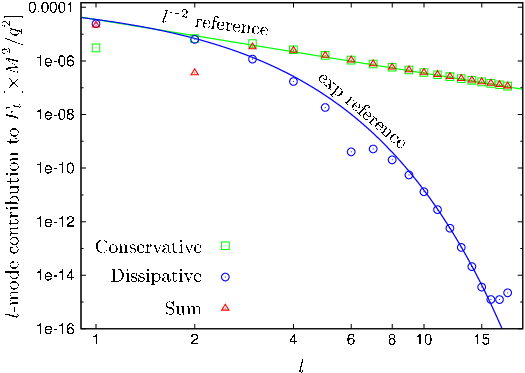}
	\hskip1cm
	\includegraphics[width=80mm]{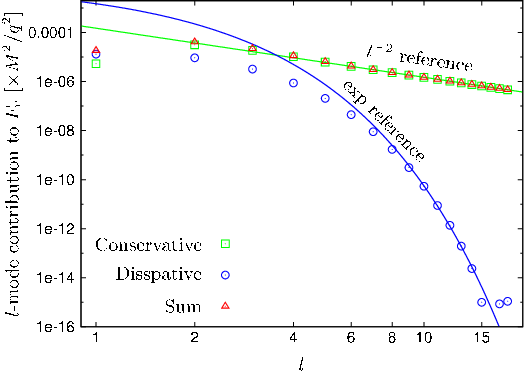}
	\end{center}
	\caption{ Behavior of the $l$-mode contributions to the mode sum for $F_t$ (left panel) and $F_r$ (right panel), shown separately for the conservative and dissipative pieces, as well as for their sum. Here we have set $(a,p,e)=(0.5M,10M,0.2)$ and the SSF is calculated at $\chi=\pi/2$. The straight solid line is a reference line $\propto l^{-2}$. From standard mode-sum theory we expect the $l$-mode contributions to the full SSF and to the conservative component of the SSF to fall off as $l^{-2}$ for large $l$. The curved solid line is an exponential reference line. We expect the $l$-mode contributions to the dissipative component of the SSF to decay exponentially with $l$.  Numerical round-off error dominates the behavior of the dissipative modes for $l \gtrsim 15$}\label{fig:normal-point}
\end{figure}

We now outline the explicit steps in our numerical calculation.
\begin{itemize}
	\item{\textit{Orbital parameters}. For given spin parameter $a$, orbital eccentricity $e$ and semi-latus rectum $p$, calculate all the necessary kinematical entities associated with the geodesic orbit ($\en, \ang, \Omega_r, \Omega_\varphi, T_r$, etc.) using the formulae given in section \ref{sec:setup}.}

	\item{\textit{Boundary conditions}. For a given $\ell m n$ mode calculate the boundary conditions using Eqs.\ \eqref{eq:inf-bc} and \eqref{eq:horiz-bc} with $c_0^{\pm} = 1$. For both boundaries we choose $k_\text{max}$ such that the relative magnitude of the $k_\text{max}+1$ term drops below a given threshold which we take to be $10^{-12}$. In order to solve the radial equation \eqref{eq:rsradialeqn} we first need to numerically invert Eq.\ \eqref{eq:tortoise-explicit} to get $r(r_*)$. Machine accuracy places a limit on the smallest $r_*$ that the inverter can resolve and we take this value to be the inner boundary of our numerical domain $r_{*\text{in}}$. This value depends on $a$. For $a=0$ we find $r_{*\text{in}} \simeq -65M$ and for $a=\pm 0.998M$ we find $r_{*\text{in}}\simeq-262M$.

There is a fair amount of freedom in the choice of the location of the outer boundary $r_{*\text{out}}$. Calculating the boundary condition using Eq.\ \eqref{eq:inf-bc} is computationally cheap in comparison to solving the radial equation, and therefore it is worth trying to place the outer boundary as close in as possible. In practice, for each $\ell m n$ we initially attempt to compute the boundary condition with $r_{*\text{out}} = 1000M$. For most modes this is adequate but for a few modes we find that the terms in the series start to grow after initially decreasing. If we detect this behavior we move $r_{*\text{out}}$ out by $5000M$ and try again. We repeat this procedure (at each stage moving the boundary out by $5000M$) until the series converges to within the required accuracy. 
 }
	\item{\textit{Homogeneous solutions}. Using the boundary conditions as determined above we numerically solve the radial equation \eqref{eq:rsradialeqn} for $\psi^\pm_{\ell m\omega}$ using the Runge-Kutta Prince-Dormand (8,9) [\texttt{gsl\_odeiv\_step\_rk8pd}] method from the Gnu Scientific Library (GSL) \cite{GSL}. As we are using the method of EHS we solve for the outer field $\psi^+_{\ell m \omega}$ between $r_{*\text{out}}$ and $r_{*\text{min}}$ and for the inner field $\psi^-_{\ell m \omega}$ from $r_{*\text{in}}$ to $r_{*\text{max}}$. We store the value of the fields $\psi^\pm_{\ell m\omega}$ and their $r_*$ derivatives at a dense sample of points between $r_\text{min}$ and $r_\text{max}$, equally spaced in $\chi$.  }

	\item{\textit{Inhomogeneous solutions}. The next step is to compute the scaling coefficients $C_{lm\omega}^\pm$ using Eq.\ \eqref{eq:scaling-coefficients}. In practice we calculate the integral in \eqref{eq:scaling-coefficients} in terms of $\chi$ rather than $t$, using  Eq.\ \eqref{eq:t_of_chi} of Appendix \ref{sec:equatorial-geodesics} to convert between the two. The integration is performed numerically using the standard QAG adaptive integrator [\texttt{gsl\_integration\_qag}] (with the 61 point Gauss-Kronrod rule) from the GSL. The integrator automatically requests the values of the integrand between $r_\text{min}$ and $r_\text{max}$ (or equivalently between $\chi=0$ and $\chi=\pi$) required to perform the integral to within a set relative accuracy of $10^{-12}$. The values of $\psi^\pm_{\ell m\omega}$ requested by the integrator are generated by locating the nearest of our sampling points stored between $r_\text{min}$ and $r_\text{max}$ and using the value of the field and its derivative stored in the previous step as input to the Runge-Kutta algorithm in order to calculate the value of the field at the requested point.}

	\item{\textit{Determining $n_\text{max}$}. Although the sum in Eq.\ \eqref{eq:EHS} is formally over all $n$, in practice we of course sum over a finite number of modes, with $|n| \leq n_\text{max}$. In our code we calculate the $n=0$ mode, then the $n=-1,1,-2,2\dots$ modes (in this order), until the relative $n$-mode contributions to both the field and its $r$ derivative drop below a given threshold (which we take to be $10^{-11}$). Typically we find that the contribution from the negative $n$ modes drops below the threshold before the equivalent positive $n$ mode does, especially for higher eccentricities (a similar behavior has been observed by Hopper and Evans \cite{Hopper-Evans}).  	}

	\item{\textit{Spherical-harmonic projection}. Once we have computed all the required $\ell m n$ modes up to some $\ell=\ell_\text{max}$ (see below) we construct the extended spherical-harmonic $\tilde{\phi}_{lm}^\pm$ modes using Eq.\ \eqref{eq:EHS}. The actual time-domain $l$-mode is then constructed using Eq.\ \eqref{eq:actual-time-domain-function}, and the $l$-mode contribution to the full force is obtained via Eq.\ \eqref{eq:field-spherical-mode}. When $a \neq 0 $ the coupling between the spheroidal and spherical harmonics implies that some of the spherical-harmonic modes with $l \le \ell_\text{max}$ will have significant contributions from uncomputed spheroidal modes with $\ell > \ell_\text{max}$. We denote by $l_\text{max}$ the highest spherical-harmonic mode for which the uncomputed spheroidal modes have a total relative contribution $<10^{-12}$. }

	\item{\textit{Large-$l$ tail contribution}. To calculate the $t,r,\varphi$ components of the SSF it is convenient to split them into conservative and dissipative pieces using Eqs.\ \eqref{eq:diss-cons}. Regularization of the two pieces is then done using Eqs.\ \eqref{eq:cons-regularization} and \eqref{eq:diss-regularization}. The dissipative component requires no regularization and it converges exponentially fast, whereas the conservative piece converges like $\propto l^{-1}$---see Fig. \ref{fig:normal-point} for an example. For a typical $\ell_\text{max}=25$ the dissipative component is computed to a high degree of accuracy, but the slow convergence of the conservative piece necessitates an extrapolation of the regularized modes to account for the uncalculated large-$l$ tail. Our method for calculating the large-$l$ contribution is presented in section IV. C. of Paper I and we implement it here in the same form. We typically find the contribution from the first few $l$ modes to be opposite in sign compared to that of the remaining modes, which sometimes results in that the contribution from the extrapolated $l > 25$ tail becomes as large as that of the calculated $l\le25$ part of the sum. The error from extrapolating the high-$l$ tail of the mode sum is by far the largest source of numerical error in the calculation of the conservative SSF.

Once the conservative and dissipative pieces of the SSF have been obtained, the full SSF is calculated by simply adding the two together. Constructing the full SSF this way (i.e., conservative and dissipative pieces in separate) has practical advantages. For example, the behavior of the large-$l$ contributions to the full component $F_t$ near the orbital turning points transitions from an exponential decay (at smaller $l$) to an $\propto l^{-2}$ fall off (at larger $l$), as demonstrated in Fig. \ref{fig:transition}. It is then difficult to estimate the tail contribution to the full SSF component $F_t$ directly. However, when separated, the conservative and dissipative pieces of $F_t$ exhibit clear power-law and exponential behaviors, respectively.}
\end{itemize}

\begin{figure}
	\begin{center}
	\includegraphics[width=80mm]{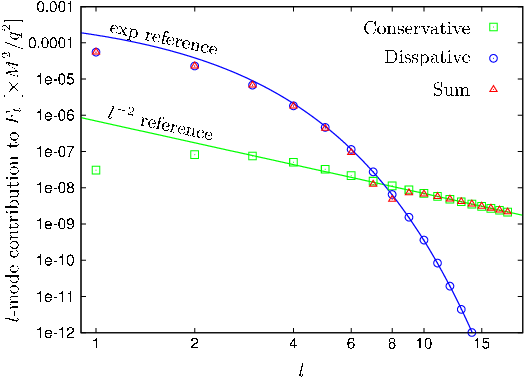}
	\hskip1cm
	\includegraphics[width=80mm]{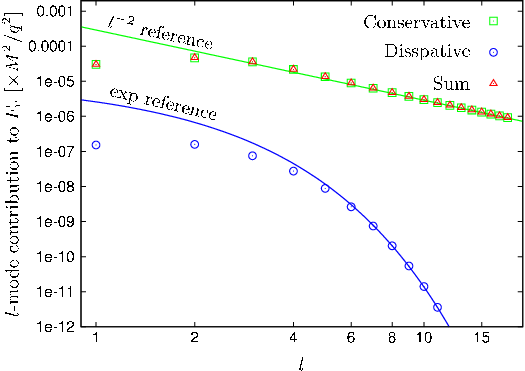}
	\end{center}
	\caption{ Same as in Figure \ref{fig:normal-point}, this time showing the $l$-mode contributions to the SSF at $\chi=0.010472$ ($=2\pi/300$), i.e., very close to periastron at $r=r_\text{min}$. Near the orbital turning points the contribution to the full $F_t$ (triangles, left panel) transitions from an exponential decay to an $\propto l^{-2}$ fall off. This transition (in this example around $l\approx8$) makes extrapolating the large-$l$ tail of the full SSF difficult in this portion of the orbit. This problem is completely avoided if we separate the SSF into conservative and dissipative pieces, calculate the large-$l$ tail contribution to the conservative piece, and then add the two pieces together to recover the full SSF. Similar behaviour is observed for the $F_\varphi$ component (not shown). No transition is observed for $F_r$ (right panel) as in this case the conservative piece dominates the SSF even near the orbital turning points. }\label{fig:transition}
\end{figure}

\section{Code validation and results}\label{sec:results}

The analysis of large-$l$ behavior, demonstrated above in Figs.\ \ref{fig:normal-point} and \ref{fig:transition}, gives an important internal validation test of our numerical procedure and of the code itself. This is particularly so for the conservative piece, where the observed $l^{-2}$ fall-off of the modal contributions rely on a delicate cancellation of as many as 3 terms in the $1/l$ expansion [the $O(l)$, $O(1)$ and $O(l^{-1})$ terms]. This test obviously probes only the large-$l$ contribution to the SSF. A complementary test, which probes primarily the low-$l$ portion of the mode sum, is achieved by comparing the work done by the (dissipative piece of the) numerically computed SSF with the asymptotic fluxes of energy and angular momentum in the scalar radiation, which we may extract independently from the numerical data.

In this section we present sample results from our SSF code, and in particular discuss the above flux comparison, which we consider a strong quantitative validation test. Since our work represents a first full implementation of the method of EHS in a SF calculation, we then take some time (in subsec.\ \ref{sec:efficiency}) to examine the efficiency of our method.

\subsection{Dissipative piece of the SSF}\label{sec:fluxes}

In Paper I we derived formulae for the average flux of scalar energy radiated to null infinity and through the event horizon. The formulae we derived specialized to circular equatorial orbits but the method of derivation was quite general and is readily extended to eccentric equatorial orbits. The resulting equations are given by
\begin{eqnarray}
	\langle\dot{E}_+\rangle 	&=&	\frac{1}{4\pi}\sum_{\ell mn} \omega_{mn}^2 | \tilde{C}_{\ell m n}^+ |^2 , 	\qquad	\langle\dot{E}_-\rangle	=	\frac{M}{2\pi r_+}\sum_{\ell mn} \omega_{mn}(\omega_{mn} - m\Omega_+) |\tilde{C}_{\ell mn}^-|^2		\c		\label{Edot}	\\
	\langle\dot{L}_+\rangle	&=& \frac{1}{4\pi}\sum_{\ell mn} m \omega_{mn} |\tilde{C}_{\ell mn}^+|^2, 	\qquad \langle\dot{L}_-\rangle	= \frac{M}{2\pi r_+}\sum_{\ell mn} m (\omega_{mn} - m\Omega_+) |\tilde{C}_{\ell mn}^-|^2						\c \label{Ldot}
\end{eqnarray}
where the $+$ and $-$ subscripts denote the flux radiated to infinity and down the event horizon, respectively, $\Omega_+=a/(2Mr_+)$, an overdot denotes differentiation with respect to $t$, and $\langle\cdot\rangle$ denotes a $t$-average over an orbital period $T_r$. The amplitude coefficients $\tilde{C}_{\ell mn}^\pm$ are given by 
\begin{equation}
\tilde{C}_{\ell mn}^\pm = C_{\ell mn}^\pm c_0^\pm,
\end{equation}
where, recall, $C_{\ell mn}^\pm$ are given in Eq.\ (\ref{eq:scaling-coeffs}), and $c_0^\pm$ are the leading coefficients from Eqs.\ (\ref{eq:inf-bc}) and (\ref{eq:horiz-bc}).
From the form of $\langle\dot{E}_-\rangle$ and $\langle\dot{L}_-\rangle$ it can be seen that the $\ell mn$-mode contribution to the horizon flux becomes negative whenever $\omega_{mn} < m \Omega_+$. For these {\it superradiant} modes \cite{Misner-superradiance} the scalar particle gains energy and angular momentum at the expense of the mass and rotation of the central black hole.

Using Eqs.\ \eqref{eq:orthogonal-force} and \eqref{eq:mass-change}, and noting that the conservative contribution to the averages $\langle \dot{\en} \rangle$ and $\langle \dot{\ang} \rangle$ vanishes by virtue of Eq.\ (\ref{eq:diss-cons}), we obtain the relations
\begin{eqnarray}
	\mu \langle \dot{\en} \rangle &=& - \frac{1}{T_r} \left( \int^{T_r}_0 \frac{F_t^\text{diss}(t)}{u^t(t)} dt + \en \Delta \mu \right)		\c	\label{eq:radiated-energy}\\
	\mu \langle \dot{\ang} \rangle &=&  \frac{1}{T_r} \left( \int^{T_r}_0 \frac{F_\varphi^\text{diss}(t)}{u^t(t)} dt - \ang \Delta \mu \right)	\c  \label{eq:radiated-ang-mom}
\end{eqnarray}
where we have used $u_t = -\en$ and $u_\varphi = \ang$, and where $\Delta\mu$ is the net change in the particle's rest mass over a period $T_r$. The latter is found from Eq.\ \eqref{eq:mass-change-explicit} to be identically zero,
\begin{eqnarray}
	\Delta\mu = 0,
\end{eqnarray}
since in our case $\Phi^R(\tau)$ comes back to itself after a period $T_r$.
The orbital energy and angular momentum dissipated by the SSF over a period $T_r$ should be balanced by the total energy and angular momentum radiated to infinity and down through the event horizon over that same period, i.e.,
\begin{eqnarray} 
-\mu\langle \dot{\en} \rangle &=& \langle \dot{E} \rangle_\text{total} \equiv \langle \dot{E}_+\rangle + \langle \dot{E}_-\rangle , \label{Ebalance}		\\
-\mu\langle \dot{\ang} \rangle &=& \langle \dot{L} \rangle_\text{total} \equiv \langle \dot{L}_+\rangle + \langle \dot{L}_-\rangle . \label{Lbalance}
\end{eqnarray}

We used our code to calculate both sides of the balance equations (\ref{Ebalance}) and (\ref{Lbalance}) for a variety of orbits and black holes spins. Table \ref{table:fluxes} summarizes a small sample of our results. In all cases considered we found a good agreement between the local dissipative SSF and the radiated fluxes. This comparison tests primarily the low-$l$ modal contributions to the SSF, because the amplitude of these contributions falls exponentially with $\ell$.
\begin{table}[htb]
    \begin{center}
        \begin{tabular}{|c c c || c c || c c |}
          \hline
            $a/M$ & $p/M$ 	&	$e$    	& $\langle \dot{\en} \rangle  \times\mu(M/q)^2$ & $1-\mu|\langle \dot{\en} \rangle/\langle \dot{E} \rangle_\text{total}|$	& $\langle \dot{\ang} \rangle \times \mu M/q^2$	& $1-\mu|\langle \dot{\ang} \rangle/\langle \dot{L} \rangle_\text{total}|$   \\
            \hline
			0.9	& 10	&	0.2		&	$2.6862422\times10^{-5}$					&	$-7.4\times10^{-8}$						& $-8.3593539\times10^{-4}$		&  $-7.4\times10^{-8}$			\\
			0.9	& 10	&	0.5		&	$2.4856622\times10^{-5}$					&	$9.5\times10^{-8}$						& $-6.2962019\times10^{-4}$		&  $7.2\times10^{-8}$			\\
			0	& 10	&	0.2		&	$3.213314062\times10^{-5}$					& 	$1.2\times10^{-10}$						& $-9.62608845\times10^{-4}$	&  $1.0\times10^{-10}$	 		\\
			0	& 10	&	0.5		&	$3.329332\times10^{-5}$						&	$1.6\times10^{-7}$						& $-7.844684\times10^{-4}$		&  $3.3\times10^{-7}$	 		\\
			-0.5 & 10	&	0.2		&	$3.65656098\times10^{-5}$					&	$2.8\times10^{-10}$						& $-1.06932319\times10^{-3}$	&  $2.0\times10^{-10}$			\\
			-0.5 & 10	&	0.5		&	$4.33567\times10^{-5}$						&	$3.1\times10^{-6}$						& $-9.70033\times10^{-4}$		&  $2.4\times10^{-6}$			\\
			0.2	 & 6.15 & 0.4		&	$3.42797\times 10^{-4}$						&	$2.5\times10^{-6}$						& $	-3.92668\times10^{-3}$		&  $2.2\times10^{-6}$			\\
          \hline
        \end{tabular}
\caption{Orbital energy and angular momentum dissipated by the SSF and comparison with the radiated fluxes, for a variety of orbits with $p=10M$. The last row shows data for a ``zoom-whirl''-type orbit (cf.\ Fig. \ref{fig:zoom-whirl}). The average dissipation rates $\langle \dot{\en} \rangle$ and $\langle \dot{\ang} \rangle$ (4th and 6th columns) are calculated from the local SSF using  Eqs.\ \eqref{eq:radiated-energy} and \eqref{eq:radiated-ang-mom}. The radiated energy and angular momentum $\langle \dot{E} \rangle_\text{total}$ and 
$\langle \dot{L} \rangle_\text{total}$ are extracted independently from the asymptotic fluxes using Eqs.\ (\ref{Edot}) and (\ref{Ldot}). The relative differences displayed in the 5th and the last columns verify that the balance relations (\ref{Ebalance}) and (\ref{Lbalance}) are satisfied. We believe the dominant source of residual discrepancy comes from the numerical integration in Eqs.\ \eqref{eq:radiated-energy} and \eqref{eq:radiated-ang-mom}.}
	\label{table:fluxes}
    \end{center}
\end{table}

\subsection{Sample results}\label{sec:sample-results}

\begin{figure}
	\includegraphics[width=80mm]{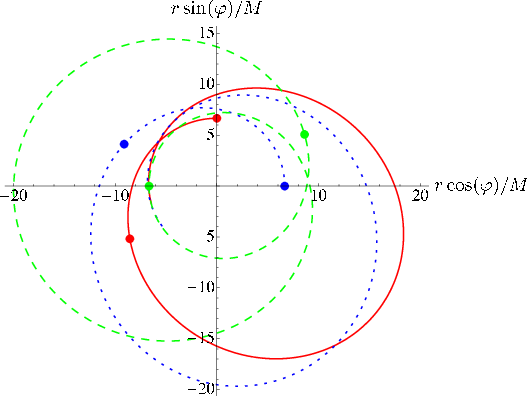}\vspace{0.5cm}\hskip0.5cm
	\includegraphics[width=80mm]{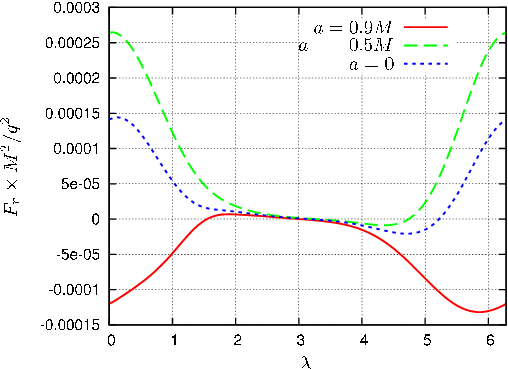}
	\includegraphics[width=80mm]{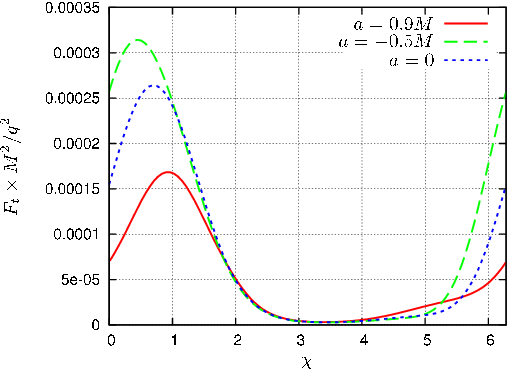}\hskip0.6cm
	\includegraphics[width=80mm]{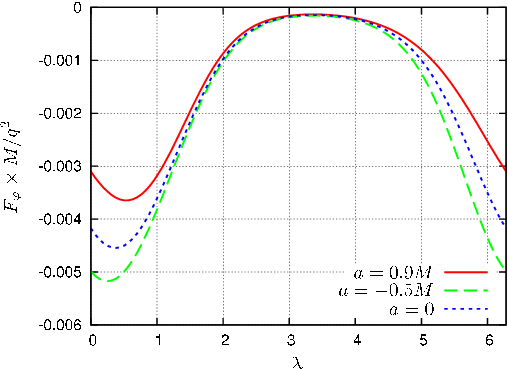}
	\caption{Sample SSF results. The top left panel shows orbits with $(p,e) = (10M, 0.5)$ in the equatorial plane, for three different black hole spins: $a=0$ (dotted, blue curve), $a=0.9M$ (solid, red curve) and $a=-0.5M$ (dashed, green curved). For each orbit we show one complete revolution, from one periastron to the next, with markers indicating the points taken for the sample data in Table \ref{table:sample-results}. The other three panels show (reading clockwise) the $F_r$, $F_\varphi$ and $F_t$ components of the SSF, for the three orbits shown in the top left panel. $\chi=0$ corresponds to a periastron passage. }\label{fig:sample-results}
\end{figure}

\begin{figure}
	\includegraphics[width=80mm]{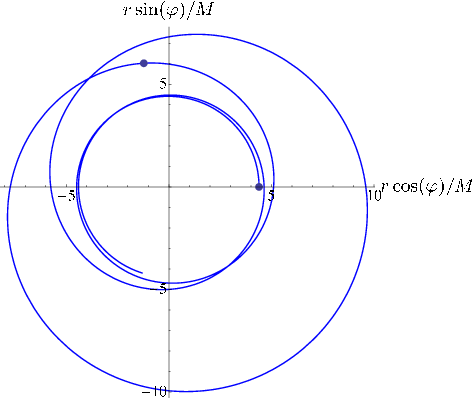}
	\includegraphics[width=80mm]{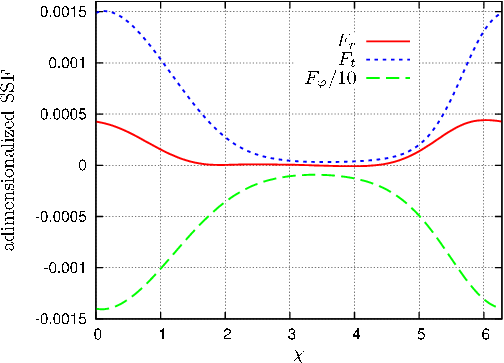}
	\caption{Shown in the left panel is a `zoom-whirl'-type orbit with parameters $(a,p,e)=(0.2M,6.15M,0.4)$. The right panel shows the corresponding components of the SSF along the orbit. $\chi=0$ is a periastron of the orbit. Markers indicate the location of the data points shown in Table \ref{table:sample-results}.}\label{fig:zoom-whirl}
\end{figure}

Using the algorithm outlined in section \ref{sec:algorithm} we have calculated the SSF for a variety of black hole spins and orbital parameters. The highest eccentricity we have been able to explore is around $e=0.7$ (see below for a discussion of limiting factors). In Fig.\ \ref{fig:sample-results} we present sample results for the SSF along a variety of orbits and in Fig.\ \ref{fig:zoom-whirl} we show an example of a `zoom-whirl'-type orbit. Table \ref{table:sample-results} displays some numerical results for the SSF for various orbits and spin values.

In the Schwarzschild case ($a=0$) it is possible to compare our results with those from the recent analysis of Ca\~nizeras \etal\,\cite{Canizares-Sopuerta:eccentric}, who used a pseudospectral algorithm formulated in the time domain. We find a good agreement---see Appendix \ref{apdx:canizares-comparison}. We have also tested the output of our code (in the $a=0$ case) against more detailed (unpublished) data from a time-domain code by Haas \cite{Haas}.

\begin{table}[htb]
    \begin{center}
		\footnotesize
        \begin{tabular}{| c c c c | c c c c c c |}
          \hline
            $a/M$ & 	$p/M$	&	$e$ &	$\chi$    & $(M^2/q^2)F_t^\text{cons} $	& $(M^2/q^2)F_t^\text{diss} $	& $(M/q^2)F_\varphi^\text{cons}$& $(M/q^2)F_\varphi^\text{diss}$& $(M^2/q^2)F_r^\text{cons}$ & $(M^2/q^2)F_r^\text{diss}$   \\
            \hline
			0.9		& 10 & 	0.2		& 0				  &	0							&	$4.9986822\times10^{-5}$	&	0							& $-1.7303353\times10^{-3}$		&	$-4.176(4)\times10^{-5}$		&	0								\\
			0.9		& 10 & 	0.2		& $\pi/2$		  & $1.5694(3)\times10^{-5}$	&	$3.6334552\times10^{-5}$	&	$-2.3979(7)\times10^{-4}$	& $-1.0515349\times10^{-3}$		&	$-1.9233(9)\times10^{-5}$		&	$1.391751\times10^{-5}$			\\
			\hline
			0.9		& 10 & 	0.5		& 0				  &	0							&	$7.5738990\times10^{-5}$	&	0							& $-3.2035416\times10^{-3}$		&	$1.120(7)\times10^{-4}$			&	0								\\
			0.9		& 10 &  0.5		& $\pi/2$		  & $4.461(2)\times10^{-5}$		&	$6.0154547\times10^{-5}$	&	$-6.072(3)\times10^{-4}$	& $-1.1478358\times10^{-3}$		&	$-2.746(4)\times10^{-5}$		&	$3.116102\times10^{-5}$			\\
			\hline
			0		& 10 & 	0.2		& 0				  &	0							&	$7.0051203\times10^{-5}$	&	0							& $-2.0550050\times10^{-3}$		&	$4.051(2)\times10^{-5}$			&	0								\\
			0		& 10 &	0.2		& $\pi/2$		  & $2.0871(1)\times10^{-5}$	&	$4.1885325\times10^{-5}$	& 	$-2.5827(3)\times10^{-4}$	& $-1.2029711\times10^{-3}$ 	&  	$1.1272(3)\times10^{-5}$		&  	$8.8783391\times10^{-6}$		\\
			\hline
			0		& 10 & 	0.5		& 0				  &	0							&	$1.5516962\times10^{-5}$	&	0							& $-4.1743275\times10^{-3}$		&	$1.446(2)\times10^{-4}$			&	0								\\
			0 		& 10 &	0.5		& $\pi/2$		  & $5.6825(3)\times10^{-5}$	&	$6.5775426\times10^{-5}$	&	$-6.6652(8)\times10^{-4}$	& $-1.2989343\times10^{-3}$ 	&	$-3.06717(7)\times10^{-6}$		&	$1.7666437\times10^{-5}$		\\
			\hline
			-0.5	& 10 & 	0.2		& 0				  &	0							&	$8.8065099\times10^{-5}$	&	0							& $-2.3164172\times10^{-3}$		&	$8.548(2)\times10^{-5}$			&	0								\\
			-0.5	& 10 &  0.2		& $\pi/2$		  &	$2.0868(1)\times10^{-5}$		&	$4.4975282\times10^{-5}$	&	$-2.4458(3)\times10^{-4}$	& $-1.3172287\times10^{-3}$		&	$3.09962(3)\times10^{-5}$		&	$9.3313239\times10^{-6}$	\\
			\hline
			-0.5	& 10 & 	0.5		& 0				  &	0							&	$2.5761765\times10^{-4}$	&	0							& $-4.9889604\times10^{-3}$		&	$2.695(1)\times10^{-4}$			&	0								\\
			-0.5	& 10 &	0.5		& $\pi/2$		  & $5.4479(5)\times10^{-5}$	&	$6.2299563\times10^{-5}$	&	$-6.421(1)\times10^{-4}$	& $-1.4030678\times10^{-3}$		&	$2.0327(7)\times10^{-5}$		&	$2.1787580\times10^{-5}$		\\
          	\hline
			0.2		& 6.15 & 0.4	& 0				  & 0							&	$1.48866752\times10^{-3}$	&	0							& $-1.4008151\times10^{-2}$		&	$4.33(5)\times10^{-4}$			&	0								\\
			0.2 	& 6.15 & 0.4	& $\pi/2$		  & $2.2520(6)\times10^{-4}$	&	$3.27980552\times10^{-4}$	&	$-1.4005(8)\times10^{-3}$	& $-4.6436085\times10^{-3}$		&	$3.712(9)\times10^{-5}$			&	$-1.563318\times10^{-5}$		\\
 			\hline
        \end{tabular}
\caption{Numerical results for the dissipative and conservative pieces of the SSF for a sample of orbits. The full SSF is obtained by adding the two pieces together. The SSF is sampled at $\chi$ values corresponding to the points marked along the orbits in Figs.\ \ref{fig:sample-results} and \ref{fig:zoom-whirl}. Data for the conservative components include an estimate of the uncertainty from the large-$l$ extrapolation, which dominates the overall numerical error in these components; this is indicated by figures in brackets, showing the uncertainty in the last quoted decimal. We used the method described in Paper I to estimate this error. In the case of $F_{\alpha}^{\rm diss}$ no large-$l$ extrapolation is needed, and the accuracy is much improved; in this case we believe all figures shown are significant. The SSF data for this table was obtained with typical values of $l_\text{max}$ between 15 and 20. }
	\label{table:sample-results}
    \end{center}
\end{table}

\subsection{Computational performance}\label{sec:efficiency}

The computational burden increases moderately with $|a|$ and more rapidly with $e$. The larger the spin magnitude $|a|$ is, the stronger the coupling between spheroidal and spherical modes becomes, and the more $\hat l$ modes need be calculated for given $l_{\rm max}$. The higher the eccentricity $e$, the broader the Fourier spectrum becomes and the more $n$ modes need be calculated for each $\hat l,m$. Moreover, larger eccentricity also leads to a stronger spheroidal--spherical coupling, because the spheroidicity parameter $\sigma^2$ that determines the strength of this coupling is proportional to $\omega_{mn}^2$, which is larger for higher $n$ harmonics. Using the current version of our code we were able to explore spin parameters in the range $-0.99M\lesssim a\lesssim 0.99M$ and eccentricities in the range $0\lesssim e\lesssim 0.7$. Beyond these ranges the computational burden becomes prohibitive.

\begin{figure}
	\includegraphics[width=85mm]{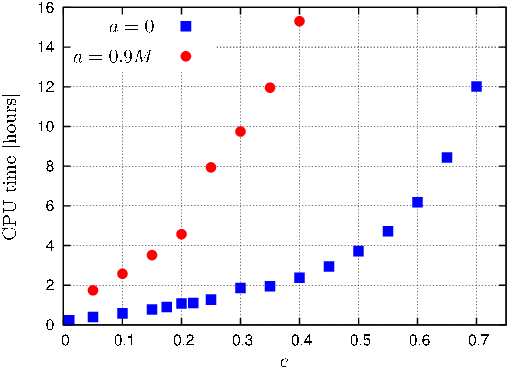}
	\caption{Computational cost. We show the total time required to calculate up to (a fiducial) $l_\text{max}= 15$, leading to fractional accuracies of order $\sim 10^{-4}$ in the SSF. In the Schwarzschild case ($a=0$) we can compute the SSF for eccentricities up to $e=0.7$ in around 12 hours. For $a=0.9M$ the calculation requires more time, primarily because the coupling between the spheroidal and spherical harmonic modes necessitates calculation of higher spheroidal harmonic modes, which is computationally expensive. At low eccentricities, our frequency-domain algorithm is vastly faster than any existing time-domain method.}\label{fig:efficiency}
\end{figure}

In Fig.\ \ref{fig:efficiency} we plot the CPU time required to compute the SSF on a standard desktop machine (dual-core, 3GHz). We used a fiducial $l_{\rm max}=15$, giving SSF fractional accuracies of order $\sim 10^{-4}$. We show results for $a=0.9M$ and, for comparison, $a=0$; results for $a=-0.9M$ are found to be similar to those for $a=0.9M$. Note that the data for $a=0$ probes the performance of the EHS method, while the Kerr results also reflect the increased computational burden due to mode coupling. 

In the Schwarzschild case ($a=0$) we find that for $e \lesssim 0.4 $ the computation time grows only $\sim$ linearly with $e$, somewhat more rapidly at higher eccentricities, and very fast for $e\gtrsim 0.6 $. Nonetheless, the required computation time for $e=0.7$ is still only around 12 hours. For comparison, an equivalent time-domain computation \cite{Hass-priv-comm} (on a similar machine and with similar accuracy standards) takes several days. At lower eccentricities our gain in speed/accuracy is very substantial. Our results for $a=0$ highlight the efficacy of the EHS method.

In the Kerr case mode coupling adds to the computational burden, and more so with growing eccentricity. As an example, for $a=0.9$ with $e=0.2$, we find that the spheroidal-harmonic $l=15$ mode has significant contributions from all tensorial-harmonic modes $8\leq \hat l\leq 22$. This results in a more rapid growth in CPU time as a function of $e$, compared to the Schwarzschild case. For spins as high as $|a|=0.9$, eccentricities greater than $\sim 0.5$ are practically beyond reach for our current code. However, for small eccentricities our algorithm is extremely efficient even at high spin.

\section{ISCO shift}\label{sec:ISCO-shift}
	
The ISCO shift due to the conservative piece of the SSF for a particle in orbit about a Schwarzschild black hole was first calculated by Diaz-Rivera \etal\ \cite{Diaz-Rivera}. More recently the ISCO shift due to the conservative piece of the {\em gravitational} SF for a similar orbital setup has also been calculated \cite{Barack-Sago-eccentric, Barack-Sago-ISCO-shift}. Here for the first time we calculate the conservative SSF correction to the innermost stable equatorial orbit (ISCEO) for a particle in orbit about a Kerr black hole. The following derivation follows closely that of Ref.\ \cite{Barack-Sago-eccentric}, but we adapt it here to Kerr spacetime.
	
\subsection{Test particle case}
	
	In this section we review the notion of the ISCEO for a test particle in Kerr geometry, ignoring all SSF effects. We begin with the radial geodesic equation for equatorial orbits in Kerr spacetime, given by \cite{Hughes}
\begin{eqnarray}\label{eq:radial-geodesic}
	\left(\diff{r_p}{\tau}\right)^2 = \frac{1}{r_p^4}\left\{ [\en (r_p^2 + a^2) - a \ang]^2 - \Delta[r_p^2 + (\ang - a \en)^2]\right\} \equiv \mathcal{R}(r_p,\en,\ang)			\p
\end{eqnarray}
Differentiating with respect to $\tau$ gives
\begin{eqnarray}\label{eq:radial-geodesic2}
	\sdiff{r_p}{\tau} =  \Feff(r_p,\en,\ang), \qquad \Feff(r_p,\en,\ang) = \frac{1}{2} \pdiff{\mathcal{R}}{r_p}		\c
\end{eqnarray}
with $\Feff$ being an effective radial acceleration.

For slightly eccentric geodesics ($e \ll 1$) we can expand the particle's radius as a function of $\tau$ in the form
\begin{eqnarray}\label{eq:radius-perturbation}
r_p(\tau) = r_0 + e r_1(\tau) + O(e^2)		\c
\end{eqnarray}
where $r_1(\tau$) is independent of $e$ and comparison with Eq.\ \eqref{eq:r_p} allows us to identify $r_0=p$. In this equation and throughout this section we use a subscript `0' to denote the circular-orbit value ($e=0$) and a subscript `1'  to denote the $O(e)$ perturbation in the quantity's value, holding $r_0$ fixed. Substituting Eq.\ \eqref{eq:radius-perturbation} into Eq.\ \eqref{eq:radial-geodesic2} and reading off the $O(e)$ terms we find
\begin{eqnarray}
 \sdiff{r_1}{\tau} =  \left.\pdiff{\Feff(r_p,\en,\ang)}{r_p}\right|_{e=0}  r_1		\c
\end{eqnarray}
where we note $\en_1 = \ang_1 = 0$ by virtue of the quadratic dependence of $\en$ and $\ang$ on $e$ [recall Eqs.\ \eqref{eq:energy} and \eqref{eq:ang-mom}, replacing $p$ with $r_0$]. Hence the $O(e)$ radial motion is a simple harmonic oscillator,
\begin{eqnarray}\label{eq:r1-harmonic}
\sdiff{r_1}{\tau} = -\omega_r^2 r_1		\c
\end{eqnarray}
with frequency
\begin{eqnarray}\label{eq:omega_r^2}
  \omega_r^2 = \left.-\pdiff{\Feff(r_p,\en,\ang)}{r_p}\right|_{e=0} = \frac{M[r_0(r_0-6M) + 8 a v r_0 - 3a^2]}{r_0^4(r_0-3M + 2av)}			\c
\end{eqnarray}
where we have denoted
\begin{eqnarray}
	v \equiv \sqrt{M/r_0} \p
\end{eqnarray}
Assuming a periapsis passage at $\tau=0$ we obtain $r_1 = -r_0\cos(\omega_r\tau)$ and hence 
\begin{eqnarray}
	r_p(\tau) = r_0(1 - e\cos\omega_r\tau) + O(e^2)	\p
\end{eqnarray}

The location of the ISCEO is defined to be the radius $r_\isco$ for which $\omega_r(r_\isco) = 0$. Solving the quartic in Eq.\ \eqref{eq:omega_r^2}, and defining $\tilde{a}=a/M$, one finds 
\begin{eqnarray}\label{ris}
	r_\isco &=& M\{ 3 + \kappa - \text{sign}(a) [(3-\gamma)(3+\gamma +2\kappa)]^{1/2} \}	\c			
\end{eqnarray}
where
\begin{eqnarray}
	\gamma		&\equiv& 1 + (1 - \tilde{a}^2)^{1/3} [(1+\tilde{a})^{1/3} + (1 - \tilde{a})^{1/3} )]	\c	\\
	\kappa		&\equiv& (3\tilde{a}^2 + \gamma^2)^{1/2}		\c
\end{eqnarray}
which to the best of our knowledge first appeared in Ref.\ \cite{Bardeen}. For the Schwarzschild and the extremal ($|a|=M$) prograde/retrograde cases the unperturbed ISCEO is located at $6M$, $1M$ and $9M$ respectively. The azimuthal orbital frequency of a test particle at the ISCEO is given by \cite{Hughes}
\begin{eqnarray}
	\Omega_{\varphi{\rm is}} = \frac{M^{1/2}}{r_\isco^{3/2} + aM^{1/2}}			\p
\end{eqnarray}
Recall our convention $\Omega_\varphi>0$, with $a>0$ for prograde orbits and $a<0$ for retrograde orbits (this is different from the convention of Ref.\ \cite{Hughes}). 

\subsection{SSF correction to the ISCEO}

Let us now derive the conservative SSF correction to the ISCEO location. The perturbed equations of motion including conservative-only SSF effects are
\begin{eqnarray}\label{eq:eqs-of-motion}
\diff{\pen}{\pert{\tau}} = -\mu^{-1} F_t^{\perp(\cons)}, \qquad \diff{\pang}{\pert{\tau}} = \mu^{-1} F^{\perp(\cons)}_\varphi	\c	\\
\sdiff{\pert{r}_p}{\pert{\tau}} = \Feff(\bar r_p,\pen,\pang) + \mu^{-1} F^r_{\perp(\cons)}		\c 		\label{eq:radial-sf}
\end{eqnarray}
where hereafter we denote perturbed quantities by an overbar, and we define $\pen\equiv -\bar u_t$ and $\pang\equiv \bar u_\varphi$ (no longer necessarily conserved along the orbit). We use the sub/superscript $\perp\!\!(\cons)$ to denote the conservative piece of the SSF perpendicular to the particle's 4-velocity [see Eq.\ \eqref{eq:orthogonal-force} and section \ref{sec:cons-diss}]. 

We assume that the radius $\pert{r}_p(\tau)$ of the SSF-perturbed slightly-eccentric orbit can again be formally expanded about a circular orbit of radius $r_0$, 
\begin{eqnarray}\label{eq:pert-r_p}
\pert{r}_p(\tau) = r_0 + e \pert{r}_1 (\tau) + O(e^2)		\c
\end{eqnarray}
where $\pert{r}_1$ depends of $r_0$ but not on $e$. We similarly expand
\begin{eqnarray}\label{eq:penpang}
\pen=\pen_0+e\pen_1(\tau)+O(e^2), \quad\quad \pang=\pang_0+e\pang_1(\tau)+O(e^2), 
\end{eqnarray} 
where $\pen_0$ and $\pang_0$ are the SSF-perturbed values of $\en_0$ and $\ang_0$ along the circular orbit of radius $r_0$.
To find $\pang_0$ and $\pen_0$ we simultaneously solve $d\pert{r}/d\pert{\tau} = 0 $ and $d^2\pert{r}/d\pert{\tau}^2 = 0$ [hence $\mathcal{R}(\bar r_p,\pen_0,\pang_0)=0$ with $\partial\mathcal{R}/\partial\bar r_p(\bar r_p,\pen_0,\pang_0)=0$]. This gives
\begin{eqnarray}
	\pen_0 	&=&	 (1-3v^2 + 2 \tilde{a} v^3)^{-1/2} \left [ 1 - 2v^2 + \tilde{a} v^3 - \frac{r_0}{2\mu}F^r_{\perp0} \right]			\c		\label{eq:pert-energy}\\
   	\pang_0	&=&	r_0(1-3v^2 + 2 \tilde{a} v^3)^{-1/2} \left[ v(1-2\tilde{a} v^3+\tilde{a}^2v^4)  - \frac{r_0(1+\tilde{a}v^3)}{2v\mu}F^r_{\perp0}\right]\ , \label{eq:pert-ang}
\end{eqnarray} 
where $F^r_{\perp0}$ is the circular-orbit value of $F^r_{\perp(\cons)}$ (note the $r$ component of the SSF is purely conservative along a circular orbit, so the label `cons' becomes redundant in this case).

The $O(e)$ part of Eq.\ \eqref{eq:radial-sf} now takes the form
\begin{eqnarray}\label{eq:pert_r_p-2nd_deriv}
\sdiff{\pert{r}_1}{\tau} = -\pomegar^2 \pert{r}_1		\c
\end{eqnarray}
where
\begin{eqnarray}\label{eq:pomega}
	\pomegar^2 &=& -\frac{d}{d\pert{r}_p} \left[ \Feff(\pert{r}_p,\pen,\pang) + \mu^{-1} F^r_{\perp(\cons)} \right]_{\pert{r} = r_0}			\ .
\end{eqnarray}
Here $\pen$, $\pang$ and $F^r_{\perp(\cons)}$ are thought of as functions of $\pert{r}_p$ along the orbit (for given $r_0,e$), and the $\pert{r}_p$ derivative is taken with fixed $r_0,e$. The form (\ref{eq:pomega}) assumes that $\pen$, $\pang$ and $F^r_{\perp(\cons)}$ depend explicitly on $e$ (when $r_0$ and $\pert{r}_p$ are held fixed) only through $e^2$. That this is true in the Schwarzschild case was shown in Ref.\ \cite{Barack-Sago-ISCO-shift} based on a simple symmetry argument, and the same argument carries over to the Kerr case. The perturbed radius is thus a simple harmonic oscillator in $\tau$ with frequency $\pomegar$ (for $\pomegar^2 >0$), and choosing $t=0$ at periastron passage we have
\begin{eqnarray}\label{eq:pert_r_p_explicit}
	\pert{r}_p(t) = r_0(1-e \cos\pomegar \pert{\tau})		\p
\end{eqnarray}

Using $\pert{r}_1 = -r_0 \cos \pomegar \pert{\tau}$ and recalling Eq.\ \eqref{eq:eqs-of-motion}, we may now write
\begin{eqnarray}
F^r_{\perp(\cons)} 		&=&	F^r_{\perp0} + e F^r_{\perp1} \cos \pomegar\pert{\tau} + O(e^2)		\c		\label{eq:Fr_cons_pert}		\\
F_\varphi^{\perp(\cons)} 	&=& e \pomegar F_\varphi^{\perp1} \sin \pomegar \pert{\tau} + O(e^2)	\c		\label{eq:Fphi_cons_pert}	\\
F_t^{\perp(\cons)}		&=& e \pomegar F_t^{\perp1} \sin \pomegar \pert{\tau} + O(e^2)		\c		\label{eq:Ft_cons_pert}
\end{eqnarray}
where we have defined
\begin{eqnarray}
F^r_{\perp1} 		\equiv -r_0 \left.\diff{F^r_{\perp(\cons)}}{\pert{r}_p}\right|_{\pert{r}_p=r_0}	\c 	\qquad	F_{\varphi}^{\perp1}	\equiv \mu r_0 \left. \diff{\pang}{\pert{r}_p}\right|_{\pert{r}_p=r_0}	\c \qquad	F_{t}^{\perp1}    \equiv -\mu r_0 \left. \diff{\pen}{\pert{r}_p}\right|_{\pert{r}_p=r_0}	\p 		\label{eq:Fr1-Ft1-Fphi1}
\end{eqnarray}
Then, using these definitions in Eq.\ \eqref{eq:pomega} and substituting for $\pen_0$ and $\pang_0$ from Eqs.\ \eqref{eq:pert-energy} and \eqref{eq:pert-ang}, we obtain
\begin{eqnarray}\hskip-1cm \label{omegabar2}
\pomegar^2 	&=& \omega_r^2 + \frac{F^r_{\perp1}}{r_0\mu} -  \frac{3-12v^2+9\tilde{a}v^3}{r_0(1-3v^2+2\tilde{a}v^3)} \frac{F^r_{\perp0}}{\mu} - 2 \frac{aM+a^2v+r_0(r_0-3M)v}{r_0^{9/2}\sqrt{r_0-3M +2av}} \frac{F_\varphi^{\perp1}}{\mu}	\label{eq:pert-omega_squared}		 \\&\,& - 2a\frac{a(r_0+M) + a^2 v - 3Mr_0v} {r_0^{9/2} \sqrt{r_0-3M+2av} } \frac{F_t^{\perp1}}{\mu}\c	\nonumber				\\
			&\equiv& \omega_r^2 + \Delta\omega_r^2(r_0,a)	\nonumber	\c
\end{eqnarray}
where $\Delta\omega_r^2(r_0,a)$ denotes the terms proportional to the SSF and we have neglected terms quadratic in the SSF. 

Near the ISCEO, the unperturbed frequency (squared) may be expanded in the form
\begin{eqnarray}\label{eq:delta-omega_squared}
\omega_r^2(r_0,a) = A(a) (r_0 - r_\isco) + O(r_0 - r_\isco)^2 \ ,
\end{eqnarray}
where, recall, $r_\isco$ denotes the location of the unperturbed ISCEO, given in Eq.\ (\ref{ris}), and 
\begin{eqnarray}\label{eq:A}
A(a) = \left.\pdiff{\omega_r^2}{r_0}\right|_{r_0=r_\isco} = -3M \frac{r_\isco^3 + M^2 r_\isco(18-5\tilde{a}^2 - 38\tilde{a}v_\isco) - 7\tilde{a}^2M^3 (\tilde{a} v_\isco - 4) + M r_\isco^2(13\tilde{a}v_\isco - 10)}{r_\isco^5\left[r_\isco+M(2\tilde{a}v_\isco -3)\right]^2}\c
\end{eqnarray}
with $v_\isco\equiv\sqrt{M/r_\isco}$. By definition, $\pomegar^2$ vanishes at the (shifted) location of the ISCEO: $\bar\omega^2(r_0=\bar r_{\rm is})=0$. By substituting Eq.\ \eqref{eq:delta-omega_squared} into Eq.\ \eqref{omegabar2}, setting $r_0=\bar r_{\isco}$ and $\pomegar = 0$, and solving for $\bar r_{\rm is}$ at linear order in the SSF, we find the SSF-induced shift in  ISCEO radius to be [through $O(q^2)$]
\begin{eqnarray}\label{eq:delta-r-isco}
\Delta r_\isco \equiv \pert{r}_\isco - r_\isco = -\frac{\Delta\omega_r^2(r_\isco,a)}{A(a)}				\p
\end{eqnarray}
Note on the right-hand side of Eq.\ \eqref{eq:delta-r-isco} we have substituted $r_\isco$ for $\pert{r}_\isco$ as this term is already of $O(q^2)$. When $a=0$, Eq.\ \eqref{eq:delta-r-isco} reduces (upon replacing the SSF components with the gravitational SF components) to the ISCO shift formula found in Ref.\ \cite{Barack-Sago-eccentric}, namely
\begin{eqnarray}\label{eq:schwarz-delta-r-isco}
\Delta r_\isco(a=0) = (M^2/\mu)\left(216 F^r_{\perp0\isco} - 108 F^r_{\perp1\isco} + \sqrt{3}M^{-2} F^{\perp1}_{\varphi\isco}\right)		\c
\end{eqnarray}
where the `$\isco$' subscript denotes a quantity's value at the unperturbed ISCO.

We next consider the conservative SSF shift in the azimuthal frequency. The frequency associated with the perturbed circular orbit of radius $\bar r=r_0$ is given by
\begin{eqnarray}\label{eq:pOmegaphi}
	\pert{\Omega}_{\varphi}	=
	 \diff{\pert{\varphi}_p}{\pert{t}} = \frac{d\pert{\varphi}_p/d\pert{\tau}}{d\pert{t}/d\pert{\tau}} = \frac{g^{\varphi\varphi}_0 \pang_0-g^{\varphi t}_0 \pen_0}{g^{t\varphi}_0 \pang_0 - g^{tt}_0 \pen_0}	\c
\end{eqnarray}
where $g_0^{\alpha\beta}$ are the background metric functions evaluated on the perturbed circular orbit. Substituting for $\pen_0$ and $\pang_0$ from Eqs.\ \eqref{eq:pert-energy} and \eqref{eq:pert-ang}, taking $r_0=r_\isco+\Delta r_\isco$ and keeping only terms through $O(q^2)$, we find the relative frequency shift at the ISCO to be given by
\begin{eqnarray}
\frac{\Delta{\Omega}_{\varphi\isco} }{\Omega_{\varphi\isco}} \equiv
\frac{\pert{\Omega}_{\varphi\isco}- \Omega_{\varphi\isco}}{\Omega_{\varphi\isco}} =
-\frac{3\Delta r_\isco}{2(r_\isco+a v_\isco)} - \frac{r_\isco^4(r_\isco-3M+2av_\isco)\mu^{-1}F^r_{\perp0\isco}}{2M(r_\isco+av_\isco)(r^2_\isco-2Mr_\isco+a^2)}	\p	
\label{eq:freq_shift}
\end{eqnarray}
For $a=0$ the above formula reduces (when the SSF components are replaced by the gravitational SF components) to that found in Refs.\ \cite{Barack-Sago-eccentric,Diaz-Rivera}, namely
\begin{eqnarray}\label{eq:delta-Omega_isco}
	\frac{\Delta{\Omega}_{\varphi\isco}}{\Omega_{\varphi\isco}}(a=0) =  -  \frac{\Delta r_\isco}{4M} - \frac{27M}{2\mu} F^r_{\perp0\isco} \p
\end{eqnarray}

All that remains is to rewrite the above expressions for $\Delta r_\isco$ and $\Delta{\Omega}_{\varphi\isco}$ in terms of the full Boyer-Lindquist components of the SSF (rather than the normal components $F_\alpha^\perp$). Specifically, recalling Eqs.\ (\ref{omegabar2}) and (\ref{eq:freq_shift}), we will need expressions for $F^r_{0\perp}$, $F^r_{\perp1}$, $F_\varphi^{\perp1}$ and $F_t^{\perp1}$ in terms of the quantities $F_r^{0}$, $F_r^{1}$, $F_\varphi^{1}$ and $F_t^{1}$ arising, in analogy with Eqs.\ \eqref{eq:Fr_cons_pert}-\eqref{eq:Ft_cons_pert}, from the formal $e$-expansion of the full conservative SSF:
\begin{eqnarray}
	F_r^{\cons} 		&=&	F_r^{0} + e F_r^{1} \cos \pomegar\pert{\tau} + O(e^2)		\c		\label{eq:Fr_cons_pert_full}		\\
	F_\varphi^{\cons} 		&=& e \pomegar F_\varphi^{1} \sin \pomegar \pert{\tau} + O(e^2)	\c		\label{eq:Fphi_cons_pert_full}		\\
	F_t^{\cons}			&=& e \pomegar F_t^{1} \sin \pomegar \pert{\tau} + O(e^2)		\p	\label{eq:Ft_cons_pert_full}
\end{eqnarray}
(We prefer to work with the covariant components of the SSF, which are the ones returned by our code.)
Starting with the radial component, we write 
\begin{eqnarray}\label{eq:Fr_cons_projected}
F^r_{\perp(\cons)} = (g^{r\beta} + u^r u^\beta)F_\beta^\cons \c
\end{eqnarray}
and formally expand both sides of the equation in $e$ using Eqs.\ (\ref{eq:Fr_cons_pert}) and (\ref{eq:Fr_cons_pert_full})--(\ref{eq:Ft_cons_pert_full}), noticing $u^r u^\beta F^\cons_\beta=O(e^2)$. Comparing the $O(e^0)$ and $O(e^1)$ terms on either side then readily gives  
\begin{eqnarray}
	F^r_{\perp0} &=& g^{rr}_0 F^0_r			\c													\label{eq:Fr0_perp}	\\
	F^r_{\perp1} &=& g^{rr}_0 F^1_r - r_0 \left.\diff{g^{rr}}{r}\right|_{r=r_0} F^0_r	\p		\label{eq:Fr1_perp}
\end{eqnarray}
For the $t$ and $\varphi$ components we similarly obtain 
\begin{eqnarray}
	F^{\perp1}_{\varphi} 	&=& 	(1+\ang^2_0 g^{\varphi\varphi}_0 - \en_0\ang_0 g^{t\varphi}_0)F_\varphi^1 + (\ang_0^2 g^{t\varphi}_0 - \en_0\ang_0 g^{tt}_0) F_t^1 + \ang_0 r_0 F_r^0 \c	\label{eq:Fphi1_perp}		\\
	F^{\perp1}_t 		&=&		(1+\en^2_0 g^{tt}_0 - \en_0\ang_0 g^{t\varphi}_0) F_t^1 + (\en^2_0 g^{t\varphi}_0 - \en_0\ang_0 g^{\varphi\varphi}_0)F_\varphi^1 - \en_0 r_0 F_r^0 		\ ,	\label{eq:Ft1_perp}
\end{eqnarray}
where we have also used $u^r = er_0\pomegar \sin\pomegar\pert{\tau}$ [recall Eq.\ \eqref{eq:pert_r_p_explicit}].

The shifts in the location and frequency of the ISCEO can now be constructed from the full SSF by substituting Eqs.\ \eqref{eq:Fr0_perp}--\eqref{eq:Ft1_perp} (evaluated at the ISCEO) into Eqs.\ \eqref{eq:delta-r-isco} and \eqref{eq:freq_shift}. The resulting formulae are cumbersome so we leave them implicit. For $a=0$ the formula for the radial ISCEO shift is found to reduce to 
\begin{eqnarray}
	\Delta r_\isco(a=0) = (M^2/\mu) \left(216 F^0_{r\isco} - 72 F_{r\isco}^1 + 6\sqrt{2} F_{t\isco}^1 + \frac{4}{\sqrt{3}} F_{\varphi\isco}^1 \right)		\c
\end{eqnarray}
which is in agreement with Eqs.\ (51) of Diaz-Rivera \etal\ \cite{Diaz-Rivera}.

\subsection{Numerical results}

\begin{figure}[htb]
	\includegraphics[width=80mm]{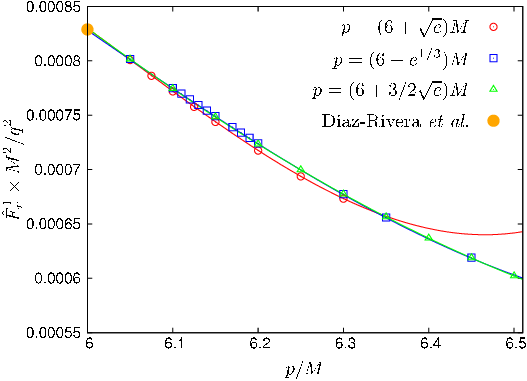}
	\includegraphics[width=80mm]{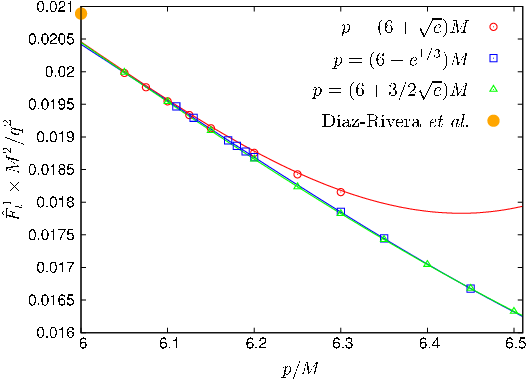}
	\includegraphics[width=80mm]{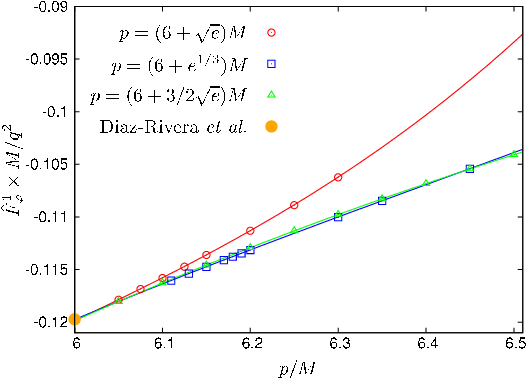}
	\includegraphics[width=80mm]{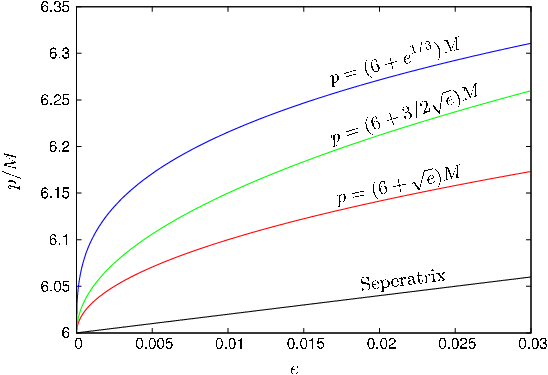}
	\caption{Calculation of $F^1_{r\isco}$, $F^{1}_{t\isco}$ and $F^{1}_{\varphi\isco}$ by extrapolation along three paths in the $e$--$p$ plane for the case of a Schwarzschild black hole ($a=0$). The sampling paths in the $e$--$p$ plane are shown in the bottom right panel. The other three panels show numerical data points for $\hat{F}_r^1$, $\hat{F}^1_\varphi$ and $\hat{F}^1_t$ as extracted from the conservative piece of the SSF using Eqs.\ (\ref{eq:Fr1-isco}) and (\ref{eq:Ft1Fphi1-isco}). Solid curves are cubic interpolations of the numerical data points, and the extrapolated values at $p=6$ (which, in theory, should not depend on the choice of curve) represent our numerical predictions for $F^1_{r\isco}$, $F^{1}_{t\isco}$ and $F^{1}_{\varphi\isco}$. The small variance in these extrapolated values serves as a rough measure of error. The thick dot on the vertical axis marks the values found by Diaz-Rivera \etal\ \cite{Diaz-Rivera}. For $F^1_{r\isco}$ and $F^{1\isco}_\varphi$ we find a close agreement with their results, but for $F^{1\isco}_t$ there is a discrepancy at a level ($\sim 2\%$) which we cannot explain. (The code used by Diaz-Rivera \etal\ cannot be retrieved to allow a careful examination of this discrepancy \cite{Detweiler:priv_comm}; we are, however, quite confident in our results given the the good agreement between the values extrapolated from the different curves.) 
	}
	\label{fig:extrapolation-in-ep-plane}
\end{figure}

\begin{figure}[htb]
	\includegraphics[width=80mm]{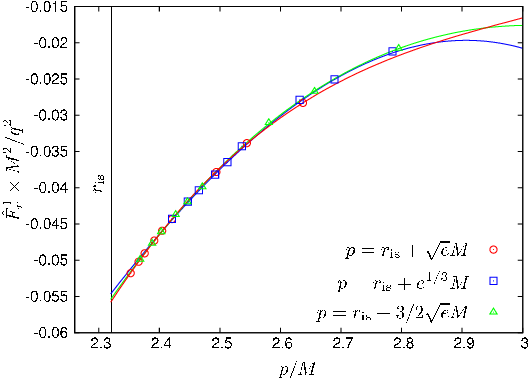}
	\includegraphics[width=80mm]{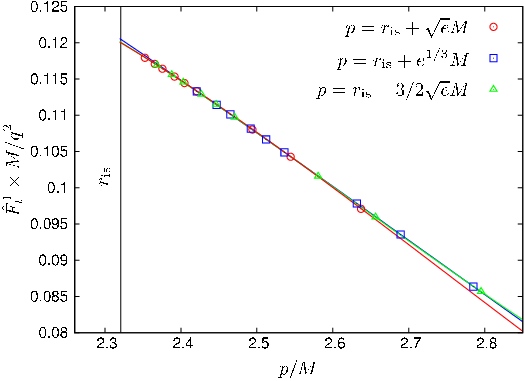}
	\caption{Same as in Fig.\ \ref{fig:extrapolation-in-ep-plane} but for $a=0.9M$. We use the paths shown on the graphs to extract the values of $F^1_{r\isco}$, $F^{1}_{t\isco}$ and $F^{1}_{\varphi\isco}$ via extrapolation to the ISCEO, whose location at $r_\isco=2.32088M$ is marked by the vertical line. (The results for $F^{1}_{\varphi\isco}$, for brevity not shown here, are qualitatively similar to those of $F^{1}_{t\isco}$.) Sample numerical values for $F^1_{r\isco}$, $F^{1}_{t\isco}$ and $F^{1}_{\varphi\isco}$, and the resulting ISCEO shifts for different spins $a$, can be found in Table \ref{table:isco-shift}. 	
}
	\label{fig:kerr-extrapolation}
\end{figure}

In order to implement Eqs.\ \eqref{eq:delta-r-isco} and \eqref{eq:freq_shift} we require the values of $F_r^0$, $F_r^{1}$, $F^{1}_t$ and $F^{1}_\varphi$, all evaluated at $r=r_\isco$. The first piece of data, $F_r^0$, is simply the radial SSF component evaluated along a circular equatorial orbit of radius $r_0=r_{\isco}$, and we can compute it with great accuracy using the circular-orbit code of Paper I. The computation of the other quantities, which are associated with a slightly eccentric orbit, is more delicate. Recalling Eqs.\ \eqref{eq:Fr_cons_pert_full}--\eqref{eq:Ft_cons_pert_full} we see that they may be extracted using 
\begin{eqnarray}
	F_{r\isco}^{1} &=& \lim_{p\rightarrow r_\isco}\lim_{e\rightarrow0} \hat{F}_r^1(p,e) \c	\qquad		\hat{F}_r^1(p,e) \equiv 2 \omega_r (e\pi)^{-1} \int^{\pi/\omega_r}_0 F_r^{\rm cons} \cos (\omega_r \tau) d\tau	\c	\label{eq:Fr1-isco}\\ 
	F^1_{\alpha\isco} &=& \lim_{p\rightarrow r_\isco}\lim_{e\rightarrow0} \hat{F}^1_\varphi(p,e) \c \qquad	\hat{F}^1_\alpha(p,e) \equiv 2(e\pi)^{-1} \int^{\pi/\omega_r}_0 F_\alpha^{\rm cons} \sin (\omega_r \tau) d\tau \c \label{eq:Ft1Fphi1-isco}
\end{eqnarray}
where $\alpha \in \{t,\varphi\}$ [we are allowed here to remove the bars off $\bar\omega_r$ and $\bar\tau$ since the quantities $F_{\alpha}$ are already $O(\mu)$]. As noted in Ref.\ \cite{Barack-Sago-eccentric}, both limits can be taken simultaneously by picking points along a suitable curve in the $e$--$p$ plane. As also discussed in Ref.\ \cite{Barack-Sago-eccentric}, for our $O(e)$-expansions to be valid we must have both $e \ll 1$ and $e \ll (p - r_\isco)/M$ along the curve (so that we keep sufficient distance from the separatrix as we approach the ISCEO). The final result should be independent of the particular path taken through the $e$--$p$ plane and we use this fact as a validation test of our numerical scheme. In practice we calculate $\hat{F}^r_1$, $\hat{F}_t^1$ and $\hat{F}_\varphi^1$ at various points along three curves given by $p=r_\isco+M\sqrt{e}$, $p=r_\isco+\frac{3}{2}M\sqrt{e}$ and $p=r_\isco + Me^{1/3}$, and then extrapolate each set of data to the ISCEO---see Figs.\ \ref{fig:extrapolation-in-ep-plane} and \ref{fig:kerr-extrapolation}. The (small) variance in the extrapolated values from the 3 curves is used as an error estimator for the $F_1$'s.

Once the $F_1$'s are at hand, we use Eqs.\ \eqref{eq:delta-r-isco} and \eqref{eq:delta-Omega_isco} to compute $\Delta r_\isco$ and $\Delta\Omega_{\varphi\isco}$ for a variety of $a$ values. The main source of error in our final results comes from the $e\to 0$ extrapolation involved in extracting the  $F_1$ functions (the error in $F_{0\isco}^r$ is relatively much smaller and can be neglected). As a rough estimator of the error in $\Delta r_\isco$ and $\Delta\Omega_{\varphi\isco}$ we use the variance in the values of these quantities when using the three different extrapolation curves mentioned above. [Notice we do not consider here the more conservative estimate obtained by adding up the contributions to the error from the various $F_1$'s (either in absolute values or in quadrature) since these are clearly correlated; we believe our estimate better represents the actual uncertainty in the final results.]


\begin{figure}
		\includegraphics[width=120mm]{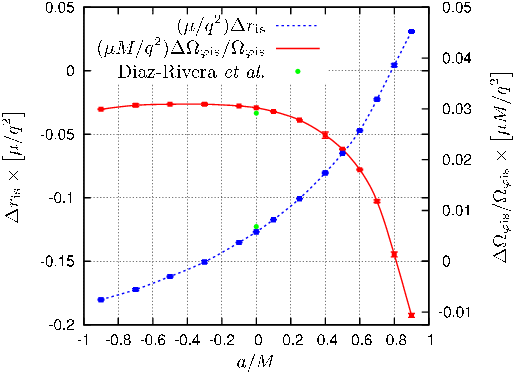}
		\caption{The conservative SSF effect upon the ISCEO location and frequency. We plot here the numerical results shown in Table \ref{table:isco-shift} as a function of the Kerr spin parameter $a$ (curves are cubic interpolations). Vertical error bars indicate the estimated numerical error. Notice in the graph that two separate scales are being used for $\Delta r_\isco$ (left-hand scale) and $\Delta\Omega_{\varphi\isco}/\Omega_{\varphi\isco}$ (right-hand scale). The radial shift is monotonically increasing with $a$ and changes sign around $a=0.8M$. The frequency shift similarly changes its sign (and turns negative) at large spin values. Note that although the change of sign in the radial and frequency shifts occur near the same spin value ($a=\sim0.8M$) the error bars on our results excludes the possiblity of the sign change occuring at the same spin value. The Schwarzschild ISCO shift results of Diaz-Rivera \etal\ \cite{Diaz-Rivera} are marked (green, thick dots) for comparison. The small discrepancy is discussed briefly in the caption of Fig.\ \ref{fig:extrapolation-in-ep-plane}.} \label{fig:isco-shift}
\end{figure}

Our results are presented in Table \ref{table:isco-shift} and Fig.\ \ref{fig:isco-shift}. We observe that $\Delta r_\isco$ increases monotonically as the black hole spin is varied from $a=-0.9M$ to $a=0.9M$, and it changes sign from negative to positive around $a=0.8M$. The relative shift in the azimuthal frequency at the ISCEO, $\Delta\Omega_{\varphi\isco}/\Omega_{\varphi\isco}$, is always positive between $a=-0.9M$ and $a=0.9M$. For retrograde orbits the relative frequency shift remains similar to that found in the Schwarzschild case ($a=0$), while for prograde orbits it decreases rapidly with increasing spin $a$. Our code is not sufficiently accurate to explore the near-extremal case, so the behavior of $\Delta\Omega_{\varphi\isco}$ (and of $\Delta r_\isco$) there remains unclear.

\begin{table}
      \begin{tabular}{|l || l l | l l l l || l l |}
          \hline
            $a/M$ 	&$r_\isco/M$& $M\Omega_{\varphi\isco}$	&	$(M^2/q^2)F_{r\isco}^0$	& $(M^2/q^2)F_{r\isco}^1 $	& $(M^2/q^2)F_{t\isco}^1 $	& $(M/q^2)F_{\varphi\isco}^1 $ 	&	 $\Delta r_\isco\left(\frac{\mu}{q^2}\right)$		& $\frac{\Delta\Omega_{\varphi\isco}}{\Omega_{\varphi\isco}}\left(\frac{\mu M}{q^2}\right)$			\\
            \hline
			0.9		& 2.32088	&	0.225442				& $-1.13367\times10^{-2}$	& 	$-0.0551(4)$			& 	0.1194(7)				& 	$-0.2814(2)$				&	0.0309(5)				&	-0.0107(3)									\\
			0.8		& 2.90664	&	0.173747				& $-3.09398\times10^{-3}$	&	$-0.0149(6)$			&	0.0822(1)				&	$-0.230(6)$					&	0.004(1)				&	0.0013(5)									\\
			0.7		& 3.39313	&	0.143879				& $-1.08845\times10^{-3}$	& 	$-4.96(9)\times10^{-3}$	& 	0.06215(7)				& 	$-0.1985(8)$				&	-0.0226(7)				&	0.0118(7)									\\
			0.6		& 3.82907	&	0.123568				& $-3.71617\times10^{-4}$	&	$-1.77(5)\times10^{-3}$	&	0.0499(2)				&	$-0.1818(7)$				&	-0.0471(2)				&	0.01802(6)									\\
			0.5		& 4.23300	&	0.108588				& $-6.92214\times10^{-5}$	& 	$-2.50(3)\times10^{-4}$	&	0.04105(2)				&	$-0.16449(2)$				&	-0.0651(6)				&	0.02203(5)									\\
			0.4		& 4.61434	&	0.096973				& $6.83682\times10^{-5}$	&	$3.32(3)\times10^{-4}$	&	0.03485(8)				&	$-0.15321(5)$				&	-0.0804(6)				&	0.0248(6)									\\
			0.25	& 5.15554	&	0.083639				& $1.31790\times10^{-4}$	& 	$7.53(3)\times10^{-4}$	& 	0.02783(1)				& 	$-0.13753(9)$				&	-0.10070(4)				&	0.02779(1)									\\
			0.1		& 5.66930	&	0.073536				& $1.68212\times10^{-4}$	&	$8.371(1)\times10^{-4}$	&	0.02295(2)				&	$-0.12610(2)$				&	-0.1172(2)				&	0.02948(4)									\\
			0.0		& 6.00000	&	0.068041				& $1.67728\times10^{-4}$	& 	$8.293(6)\times10^{-4}$	& 	0.02043(1)				& 	$-0.11983(5)$				&	-0.1268(1)				&	0.03020(3)									\\
			-0.1	& 6.32289	&	0.063294				& $1.62329\times10^{-4}$	&	$7.98(2)\times10^{-4}$	&	0.01833(2)				&	$-0.1141(2)$				&	-0.1350(3)				&	0.03056(6)									\\
			-0.3	& 6.94927	&	0.055496				& $1.31790\times10^{-4}$	& 	$7.13(2)\times10^{-4}$	& 	0.01504(5)				& 	$-0.1050(5)$				&	-0.1505(9)				&	0.0309(4)									\\
			-0.5	& 7.55458	&	0.049348				& $1.27517\times10^{-4}$	& 	$6.23(2)\times10^{-4}$	& 	0.01257(2)				& 	$-0.0970(2)$				&	-0.1620(4)				&	0.03090(8)									\\
			-0.7	& 8.14297	&	0.044372				& $1.10762\times10^{-4}$	& 	$5.36(2)\times10^{-4}$	& 	0.01068(3)				& 	$-0.0909(6)$				&	-0.1723(8)				&	0.0306(2)									\\
			-0.9	& 8.71735	&	0.040260				& $9.60700\times10^{-5}$	& 	$4.68(2)\times10^{-4}$	& 	$0.00919(2)$			& 	$-0.08492(7)$				&	-0.1802(4)				&	0.02992(7)									\\
          \hline
        \end{tabular}
\caption{The conservative SSF effect upon the ISCEO location and frequency. Each row of the table corresponds to a particular value of the Kerr spin parameter $a$: the second and third columns show the values of the unperturbed ISCO radius $r_{\isco}$ and frequency $\Omega_{\varphi\isco}$, and the fourth through seventh columns show the numerically-computed values of the SSF coefficients $F^0_{r\isco}$, $F^1_{r\isco}$, $F_{t\isco}^1$ and $F_{\varphi\isco}^1$ defined through the small-$e$ expansion in Eqs.\ (\ref{eq:Fr_cons_pert_full})--(\ref{eq:Ft_cons_pert_full}). The last two columns display the SSF-induced shift in the radius and frequency of the ISCEO, as computed using Eqs.\ \eqref{eq:delta-r-isco} and \eqref{eq:freq_shift}. Figures in brackets are estimates of the numerical error in the last displayed decimals (in the data for $F^0_{r\isco}$ all figures are significant). Note the fractional error in the $a=0.8M$ results for $\Delta r_\isco$ and $\Delta\Omega_{\varphi\isco}$ is particularly large: this is a consequence of a delicate cancellation between the various terms in Eqs.\ \eqref{eq:delta-r-isco} and \eqref{eq:delta-Omega_isco}, which also leads to the vanishing of $\Delta r_\isco$ and $\Delta\Omega_{\varphi\isco}$ at two (slightly different) spin values close to $a=0.8$. For $a=0$ Diaz-Rivera \etal\ \cite{Diaz-Rivera} obtained  $\Delta r_\isco=-0.122701 q^2/\mu$ and $\Delta\Omega_{\varphi\isco}/\Omega_{\varphi\isco} =0.0291657 q^2/(\mu M)$. The small discrepancy is discussed briefly in the caption of Fig.\ \ref{fig:extrapolation-in-ep-plane}.
		}\label{table:isco-shift}
\end{table}

\section{Variation of rest mass}\label{sec:masschange}

As discussed in Sec.\ \ref{sec:orbital-setup}, the SSF has a component tangential to the particle's worldline, which leads to the particle having a dynamically varying rest mass \cite{Quinn}. (This situation is special to our particular SSF theory; in the equivalent electromagnetic and gravitational cases the rest mass is conserved. It is possible to construct a scalar field theory where the particle's rest mass is conserved but only at the cost of making the field equation non-linear \cite{Quinn}.) Previous studies of this phenomenon in cosmological spacetimes \cite{Haas-Poisson-mass_change, Burko-Harte-Poisson} have found a range of possibilities including a periodic mass variation as well as cases where the mass dissipates entirely.

In our setup, where the motion is intrinsically periodic, the field returns to its original value after one orbital revolution and thus from Eq.\ \eqref{eq:mass-change-explicit} we see that the net change in the particle's rest mass will be zero. Furthermore, examining Eq.\ \eqref{eq:mass-change} and recalling the symmetry relations expressed in Eq.\ (\ref{eq:diss-cons}), we can see that $d\mu/d\tau$ is symmetric about the apastron and hence the rest-mass change from periastron to apastron (and visa versa) must also be zero. To within our numerical accuracy we observe this behavior in our data---see Figure \ref{fig:mass_change} below.

It is also interesting to examine how the rest mass varies along the orbit. The total rest mass change from periastron to a point with phase $\chi$ along the orbit is given by
\begin{eqnarray}
	\Delta\mu(\chi) = -\int^{\chi}_0  F_\alpha^\text{diss}(\chi) u^\alpha(\chi') \diff{t}{\chi'} d\chi'		\p
\end{eqnarray}
As illustrated in Fig.\ \ref{fig:mass_change}, the particle's rest mass initially increases (though for zoom-whirl type orbits there is a slight decreases before the main increase) but then decreases so that the particle regains its original mass by the time it reaches apastron. Past the apastron the mass continues to decrease before increasing back to the original value at periastron. We also observe that the change in mass along the orbit is only weakly dependent on the black hole spin, for fixed $(p,e)$.

In our setup the particle's rest mass is conserved over an orbital period $T_r$, but in a setup which allowed for the orbit to evolve through the action of the SSF this would no longer be the case (the field would no longer return to it's original value after one orbit). It would be an interesting project, which we do not pursue here, to consider the effect of the net mass loss on the insipral dynamics of the scalar charge.

\begin{figure}
	\begin{center}
	\includegraphics[width=80mm]{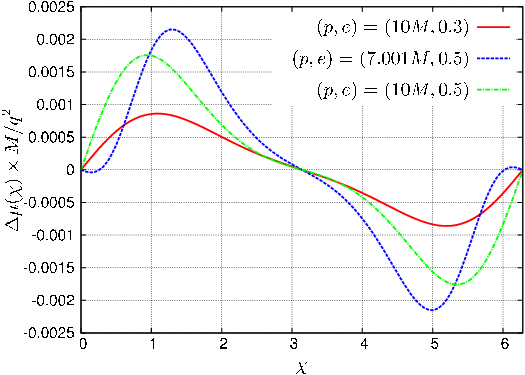}\hskip1cm
	\includegraphics[width=80mm]{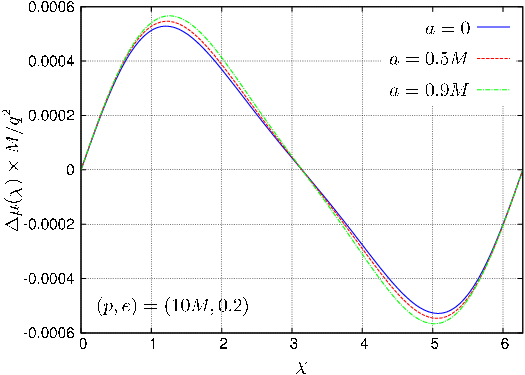}
	\end{center}
	\caption{Rest-mass change due to the SSF for a scalar charge in an eccentric equatorial orbit about a Kerr black hole. (Left panel) The variation in the rest mass, for orbits about a Schwarzschild black hole, as a function of $\chi$, is strongly dependent on the orbit's eccentricity. For $e$ and $p$ far away from the separatrix the mass initially increases and then returns to its original value at apastron before further decreasing and then once again returning to the original value at periastron. For a `zoom-whirl'-type orbit the mass is observed to decrease slightly initially. (Right panel) Results for Kerr. The change in mass is weakly dependent on the black hole spin. }\label{fig:mass_change}
\end{figure}

\section{Summary and future work}\label{sec:conclusion}

In this work we have presented a first calculation of the SSF for a particle moving along an eccentric, equatorial geodesic of a Kerr black hole. Working in the FD we have made use of the recently proposed method of EHS to overcome the difficulties that would have otherwise hampered the convergence of the frequency mode-sum (in either Schwarzschild or Kerr spacetimes). As this work represents the first complete SF calculation made using the method of EHS we have also explored the efficiency of the method and found it to perform extremely well. In the Schwarzschild case, calculating the SSF along an eccentric geodesic (at a fractional accuracy of $\lesssim 10^{-4}$, on a standard desktop computer) takes between a few minutes for small eccentricities and $\sim 12$ hours for $e=0.7$---still substantially faster than equivalent time-domain codes. Calculations in Kerr are more computationally demanding due to mode coupling (and in this case there are no time-domain codes at hand yet to allow comparison).  

In Table \ref{table:sample-results} we provide sample SSF results, which may serve as comparison data for researchers working on alternative computation methods in Kerr. We performed several validation tests to establish confidence in our results. First, we checked that, for each point along the orbit, the $l$-mode contributions to the SSF exhibit the correct large-$l$ asymptotics, as predicted by mode-sum theory. Second, we compared the rate of energy and angular momentum dissipation by the SSF with the corresponding fluxes in the scalar waves radiated down the event horizon and to infinity. Finally, for the non-rotating case we compared our results with those of Haas \cite{Haas} and Ca{\~n}izares and Sopuerta \cite{Canizares-Sopuerta:eccentric}.

We have used our numerical results to quantify two ``physical'' effects associated with the SSF in our scalar-field model. First, we calculated the shift in the ISCEO radius and frequency due to the conservative piece of the SSF, hence generalizing previous results by Diaz-Rivera \etal\ \cite{Diaz-Rivera} to the Kerr case. Our main results for the ISCEO shift are summarized in Table \ref{table:isco-shift} and plotted in Fig.\ \ref{fig:isco-shift}. We then also examined, for the first time in a black hole spacetime, the variation in the rest mass of the scalar particle resulting from the component of the SSF tangent to the four-velocity. We confirmed that the particle exchanges mass/energy with the scalar field in such a way that (if the inspiral motion is ignored) no net mass loss occurs over a complete radial period. Sample results for the mass change are presented in Fig.\ \ref{fig:mass_change}.

Several extensions of this work suggest themselves. First, one might generalize to generic, inclined orbits in Kerr geometry. We have made some progress calculating the SSF for {\em circular} inclined orbits (and will present our results elsewhere), but the challenge of generic orbits (ones that are inclined {\em and} eccentric) still lies ahead. Such generic orbits are tri-periodic, with the additional frequency coming from the longitudinal motion. This is certain to increase the computational burden and might further restrict the portion of parameter space that is tractable using our FD method.

The important extension to the gravitational case is even more challenging. It is not clear if there is a second-rank tensorial generalization of the scalar spheroidal harmonics that would facilitate a full separation of variables in the frequency domain in Kerr geometry. In the absence of such formulation one could still tackle the gravitational perturbation equations in the frequency domain, by decomposing in the standard tensorial {\em spherical} harmonics and then properly accounting for the coupling between $l$ modes that would result in Kerr. 

At least in the Schwarzschild case, however, it is clear that an EHS-based FD method is both viable and extremely computationally efficient. In this case, a standard tensor-harmonic decomposition can be employed to achieve a full separation of the gravitational perturbation equations (e.g., in Lorenz gauge), and one can proceed in a straightforward way to construct the gravitational SF using EHS in conjunction with mode-sum regularization, as in the scalar case.
This method has already been implement recently by Akcay for circular orbits \cite{Akcay-Barack}, and work to generalize it to eccentric geodesics is well under way.

\section*{ACKNOWLEDGEMENTS}

We are grateful to Roland Haas for providing us with detailed TD data for comparison. We also thank Steven Detweiler, Sam Dolan, Abraham Harte, Eric Poisson and Norichika Sago for helpful discussions. We further thank Barry Wardell and C\'esar Merl\'in for a careful reading of this work. NW's work was supported by an STFC studentship grant. LB acknowledges support from STFC through grant number PP/E001025/1.

\appendix

\section{Some expressions entering the formulation of eccentric equatorial geodesics in Kerr spacetime}\label{sec:equatorial-geodesics}

We give here a few expressions that go into the calculation of the geodesics in section \ref{sec:setup}. The azimuthal angle $\varphi_p(\chi)$ and Boyer-Lindquist time $t_p(\chi)$ along the geodesic are given by \cite{Glampedakis-Kennefick}
\begin{eqnarray}
	\varphi_p(\chi) 	&=&		\int^\chi_0 d\chi'  \frac{\tilde{V}_\varphi ( \chi',p,e)}{J(\chi',p,e) \tilde{V}_r^{1/2} (\chi',p,e)}	\label{eq:phi_of_chi}		\c		\\
	t_p(\chi)		&=&		\int^\chi_0 d\chi' \frac{\tilde{V}_t (\chi',p,e)}{J(\chi',p,e) \tilde{V}_r^{1/2} (\chi',p,e)}		\label{eq:t_of_chi}			\c
\end{eqnarray}
where
\begin{eqnarray}
	\tilde{V}_r(\chi,p,e)  			&=&		 x^2 + a^2 + 2 a x \en - \frac{2 M x^2}{p} (3+e \cos\chi)						\c		\\
	\tilde{V}_\varphi(\chi,p,e)		&=&		x  + a \en - \frac{2 M x}{p} (1+ e \cos\chi)									\c		\\
	\tilde{V}_t (\chi,p,e)			&=&		a^2 \en  -  \frac{2 a M x}{p} (1+ e \cos\chi) + \frac{\en p ^2}{(1+ e \cos\chi)^2}		\c		\\
	J(\chi,p,e)						&=&		1 - \frac{2 M}{p}(1+e \cos\chi) + \frac{a^2}{p^2}(1+e\cos\chi)^2					\p
\end{eqnarray}
The quantity $x=x(a,p,e)$ [which also enters Eqs.\ \eqref{eq:energy} and \eqref{eq:ang-mom} for the particle's energy $\en$ and angular-momentum $\ang$] is given by
\begin{eqnarray}
	x = \left[\frac{ -N -\text{sign}(a) \sqrt{N^2 - 4 F C}}{2 F} \right]^{1/2}\c
\end{eqnarray}
with
\begin{eqnarray}
	F(p,e) 	&=& \frac{1}{p^3}\left[ p^3 - 2M(3+e^2)p^2 + M^2 (3+e^2)^2 p - 4Ma^2(1-e^2)^2 \right]		\c			\\
	N(p,e) 	&=& \frac{2}{p}\left\{-Mp^2 + \left[M^2(3+e^2) - a^2\right]p - M a^2 (1+3e^2) \right\}					\c			\\
	C(p)		&=& (a^2 - Mp)^2															\p		
\end{eqnarray}

\section{Comparison with previous results in the Schwarzschild case}\label{apdx:canizares-comparison}

Ca\~nizares \etal\ \cite{Canizares-Sopuerta:eccentric} recently presented numerical results from a calculation of the SSF using a pseudospectral collocation method. Their calculation was carried out in the time domain, restricting to the Schwarzschild case ($a=0$). In Table \ref{table:canizares-comparison} we show a comparison of their results with ours, showing a good agreement.

\begin{table}[htb]
    \begin{center}
        \begin{tabular}{|c c | c | c  c | c |}
		\hline
		$p$ 		& $e$  	& SSF component 	& This work 					& Ca\~nizares \etal 			& Relative diff.	\\
		\hline
		6.3			& 0.1	& $F_t$				& $4.517994\times10^{-4}$		& $4.5171\times10^{-4}$		& 0.01\%			\\
					&		& $F_r$				& $2.1257\times10^{-4}$			& $2.1250\times10^{-4}$		& 0.03\%			\\
					&		& $F_\varphi$		& $-6.020401\times10^{-3}$		& $-6.02040\times10^{-3}$	& 0.0003\%			\\
		\hline
		6.7			& 0.3	& $F_t$				& $7.71773\times10^{-4}$		& $7.6980\times10^{-4}$		& 0.25\%			\\	
					&		& $F_r$				& $3.6322\times10^{-4}$			& $3.6339\times10^{-4}$		& 0.04\%				\\
					&		& $F_\varphi$		& $-9.04021\times10^{-3}$		& $-9.0402\times10^{-3}$	& 0.00015\%			\\
		\hline
		7.1			& 0.5	& $F_t$				& $1.22331\times10^{-3}$		& $1.2330\times10^{-3}$		& 0.015\%			\\	
					&		& $F_r$				& $5.6179\times10^{-4}$			& $5.6122\times10^{-4}$		& 0.1\%				\\
					&		& $F_\varphi$		& $-1.26857\times10^{-2}$		& $-1.2685\times10^{-2}$	& 0.0061\%			\\
		\hline
        \end{tabular}
	\caption{Comparison with Ca\~nizares \etal\ \cite{Canizares-Sopuerta:eccentric} in the {\em Schwarzschild} case ($a=0$). The SSF values are extracted at certain near-periastron points as specified in Table I of \cite{Canizares-Sopuerta:eccentric}. Ca\~nizares \etal\ do not indicate error bars on their results; for our data all figures are significant.   }
	\label{table:canizares-comparison}
    \end{center}
\end{table}


\begin{thebibliography}{36}%
\makeatletter
\providecommand \@ifxundefined [1]{%
 \@ifx{#1\undefined}
}%
\providecommand \@ifnum [1]{%
 \ifnum #1\expandafter \@firstoftwo
 \else \expandafter \@secondoftwo
 \fi
}%
\providecommand \@ifx [1]{%
 \ifx #1\expandafter \@firstoftwo
 \else \expandafter \@secondoftwo
 \fi
}%
\providecommand \natexlab [1]{#1}%
\providecommand \enquote  [1]{``#1''}%
\providecommand \bibnamefont  [1]{#1}%
\providecommand \bibfnamefont [1]{#1}%
\providecommand \citenamefont [1]{#1}%
\providecommand \href@noop [0]{\@secondoftwo}%
\providecommand \href [0]{\begingroup \@sanitize@url \@href}%
\providecommand \@href[1]{\@@startlink{#1}\@@href}%
\providecommand \@@href[1]{\endgroup#1\@@endlink}%
\providecommand \@sanitize@url [0]{\catcode `\\12\catcode `\$12\catcode
  `\&12\catcode `\#12\catcode `\^12\catcode `\_12\catcode `\%12\relax}%
\providecommand \@@startlink[1]{}%
\providecommand \@@endlink[0]{}%
\providecommand \url  [0]{\begingroup\@sanitize@url \@url }%
\providecommand \@url [1]{\endgroup\@href {#1}{\urlprefix }}%
\providecommand \urlprefix  [0]{URL }%
\providecommand \Eprint [0]{\href }%
\providecommand \doibase [0]{http://dx.doi.org/}%
\providecommand \selectlanguage [0]{\@gobble}%
\providecommand \bibinfo  [0]{\@secondoftwo}%
\providecommand \bibfield  [0]{\@secondoftwo}%
\providecommand \translation [1]{[#1]}%
\providecommand \BibitemOpen [0]{}%
\providecommand \bibitemStop [0]{}%
\providecommand \bibitemNoStop [0]{.\EOS\space}%
\providecommand \EOS [0]{\spacefactor3000\relax}%
\providecommand \BibitemShut  [1]{\csname bibitem#1\endcsname}%
\let\auto@bib@innerbib\@empty
\bibitem [{\citenamefont {{Barack}}\ and\ \citenamefont
  {{Sago}}(2010)}]{Barack-Sago-eccentric}%
  \BibitemOpen
  \bibfield  {author} {\bibinfo {author} {\bibfnamefont {L.}~\bibnamefont
  {{Barack}}}\ and\ \bibinfo {author} {\bibfnamefont {N.}~\bibnamefont
  {{Sago}}},\ }\href {\doibase 10.1103/PhysRevD.81.084021} {\bibfield
  {journal} {\bibinfo  {journal} {\prd}\ }\textbf {\bibinfo {volume} {81}},\
  \bibinfo {pages} {084021} (\bibinfo {year} {2010})},\ \Eprint
  {http://arxiv.org/abs/1002.2386} {arXiv:1002.2386} \BibitemShut {NoStop}%
\bibitem [{\citenamefont {{Warburton}}\ and\ \citenamefont
  {{Barack}}(2010)}]{Warburton-Barack}%
  \BibitemOpen
  \bibfield  {author} {\bibinfo {author} {\bibfnamefont {N.}~\bibnamefont
  {{Warburton}}}\ and\ \bibinfo {author} {\bibfnamefont {L.}~\bibnamefont
  {{Barack}}},\ }\href {\doibase 10.1103/PhysRevD.81.084039} {\bibfield
  {journal} {\bibinfo  {journal} {\prd}\ }\textbf {\bibinfo {volume} {81}},\
  \bibinfo {pages} {084039} (\bibinfo {year} {2010})},\ \Eprint
  {http://arxiv.org/abs/1003.1860} {arXiv:1003.1860} \BibitemShut {NoStop}%
\bibitem [{\citenamefont {{Barack}}\ and\ \citenamefont
  {{Ori}}(2003)}]{Barack-Ori}%
  \BibitemOpen
  \bibfield  {author} {\bibinfo {author} {\bibfnamefont {L.}~\bibnamefont
  {{Barack}}}\ and\ \bibinfo {author} {\bibfnamefont {A.}~\bibnamefont
  {{Ori}}},\ }\href {\doibase 10.1103/PhysRevLett.90.111101} {\bibfield
  {journal} {\bibinfo  {journal} {Phys. Rev. Lett.}\ }\textbf {\bibinfo
  {volume} {90}},\ \bibinfo {pages} {111101} (\bibinfo {year} {2003})},\
  \bibinfo {note} {see \cite{Barack-review} for the explicit expressions of the
  regularization parameters in terms of elliptic integrals.}\BibitemShut
  {Stop}%
\bibitem [{\citenamefont {{Barton}}\ \emph {et~al.}(2008)\citenamefont
  {{Barton}}, \citenamefont {{Lazar}}, \citenamefont {{Kennefick}},
  \citenamefont {{Khanna}},\ and\ \citenamefont {{Burko}}}]{Barton-etal}%
  \BibitemOpen
  \bibfield  {author} {\bibinfo {author} {\bibfnamefont {J.~L.}\ \bibnamefont
  {{Barton}}}, \bibinfo {author} {\bibfnamefont {D.~J.}\ \bibnamefont
  {{Lazar}}}, \bibinfo {author} {\bibfnamefont {D.}\ \bibnamefont
  {{Kennefick}}}, \bibinfo {author} {\bibfnamefont {G.}~\bibnamefont
  {{Khanna}}}, \ and\ \bibinfo {author} {\bibfnamefont {L.}\ \bibnamefont
  {{Burko}}},\ }\href {\doibase 10.1103/PhysRevD.78.064042} {\bibfield
  {journal} {\bibinfo  {journal} {\prd}\ }\textbf {\bibinfo {volume} {78}},\
  \bibinfo {pages} {064042} (\bibinfo {year} {2008})},\ \Eprint
  {http://arxiv.org/abs/0804.1075} {arXiv:0804.1075} \BibitemShut {NoStop}%
\bibitem [{\citenamefont {{Pound}}\ and\ \citenamefont
  {{Poisson}}(2008)}]{Pound-Poisson:Osculating_orbits}%
  \BibitemOpen
  \bibfield  {author} {\bibinfo {author} {\bibfnamefont {A.}~\bibnamefont
  {{Pound}}}\ and\ \bibinfo {author} {\bibfnamefont {E.}~\bibnamefont
  {{Poisson}}},\ }\href {\doibase 10.1103/PhysRevD.77.044013} {\bibfield
  {journal} {\bibinfo  {journal} {\prd}\ }\textbf {\bibinfo {volume} {77}},\
  \bibinfo {pages} {044013} (\bibinfo {year} {2008})},\ \Eprint
  {http://arxiv.org/abs/0708.3033} {arXiv:0708.3033} \BibitemShut {NoStop}%
\bibitem [{\citenamefont {Gair}\ \emph {et~al.}(2011)\citenamefont {Gair},
  \citenamefont {Flanagan}, \citenamefont {Drasco}, \citenamefont {Hinderer},\
  and\ \citenamefont {Babak}}]{Gair-etal}%
  \BibitemOpen
  \bibfield  {author} {\bibinfo {author} {\bibfnamefont {J.~R.}\ \bibnamefont
  {Gair}}, \bibinfo {author} {\bibfnamefont {E.~E.}\ \bibnamefont {Flanagan}},
  \bibinfo {author} {\bibfnamefont {S.}~\bibnamefont {Drasco}}, \bibinfo
  {author} {\bibfnamefont {T.}~\bibnamefont {Hinderer}}, \ and\ \bibinfo
  {author} {\bibfnamefont {S.}~\bibnamefont {Babak}},\ }\href {\doibase
  10.1103/PhysRevD.83.044037} {\bibfield  {journal} {\bibinfo  {journal} {Phys.
  Rev. D}\ }\textbf {\bibinfo {volume} {83}},\ \bibinfo {pages} {044037}
  (\bibinfo {year} {2011})}\BibitemShut {NoStop}%
\bibitem [{\citenamefont {{Barack}}\ and\ \citenamefont
  {{Lousto}}(2005)}]{Barack-Lousto-2005}%
  \BibitemOpen
  \bibfield  {author} {\bibinfo {author} {\bibfnamefont {L.}~\bibnamefont
  {{Barack}}}\ and\ \bibinfo {author} {\bibfnamefont {C.~O.}\ \bibnamefont
  {{Lousto}}},\ }\href {\doibase 10.1103/PhysRevD.72.104026} {\bibfield
  {journal} {\bibinfo  {journal} {\prd}\ }\textbf {\bibinfo {volume} {72}},\
  \bibinfo {pages} {104026} (\bibinfo {year} {2005})},\ \Eprint
  {http://arxiv.org/abs/arXiv:gr-qc/0510019} {arXiv:gr-qc/0510019} \BibitemShut
  {NoStop}%
\bibitem [{\citenamefont {{Barack}}\ and\ \citenamefont
  {{Sago}}(2007)}]{Barack-Sago-circular}%
  \BibitemOpen
  \bibfield  {author} {\bibinfo {author} {\bibfnamefont {L.}~\bibnamefont
  {{Barack}}}\ and\ \bibinfo {author} {\bibfnamefont {N.}~\bibnamefont
  {{Sago}}},\ }\href {\doibase 10.1103/PhysRevD.75.064021} {\bibfield
  {journal} {\bibinfo  {journal} {\prd}\ }\textbf {\bibinfo {volume} {75}},\
  \bibinfo {pages} {064021} (\bibinfo {year} {2007})},\ \Eprint
  {http://arxiv.org/abs/arXiv:gr-qc/0701069} {arXiv:gr-qc/0701069} \BibitemShut
  {NoStop}%
\bibitem [{\citenamefont {{Akcay}}(2011)}]{Akcay-Barack}%
  \BibitemOpen
  \bibfield  {author} {\bibinfo {author} {\bibfnamefont {S.}~\bibnamefont
  {{Akcay}}},\ }\href {\doibase 10.1103/PhysRevD.83.124026} {\bibfield
  {journal} {\bibinfo  {journal} {\prd}\ }\textbf {\bibinfo {volume} {83}},\
  \bibinfo {pages} {124026} (\bibinfo {year} {2011})},\ \Eprint
  {http://arxiv.org/abs/1012.5860} {arXiv:1012.5860} \BibitemShut {NoStop}%
\bibitem [{\citenamefont {{Shah}}\ \emph {et~al.}(2011)\citenamefont {{Shah}},
  \citenamefont {{Keidl}}, \citenamefont {{Friedman}}, \citenamefont {{Kim}},\
  and\ \citenamefont {{Price}}}]{Shah-etal}%
  \BibitemOpen
  \bibfield  {author} {\bibinfo {author} {\bibfnamefont {A.~G.}\ \bibnamefont
  {{Shah}}}, \bibinfo {author} {\bibfnamefont {T.~S.}\ \bibnamefont {{Keidl}}},
  \bibinfo {author} {\bibfnamefont {J.~L.}\ \bibnamefont {{Friedman}}},
  \bibinfo {author} {\bibfnamefont {D.-H.}\ \bibnamefont {{Kim}}}, \ and\
  \bibinfo {author} {\bibfnamefont {L.~R.}\ \bibnamefont {{Price}}},\ }\href
  {\doibase 10.1103/PhysRevD.83.064018} {\bibfield  {journal} {\bibinfo
  {journal} {\prd}\ }\textbf {\bibinfo {volume} {83}},\ \bibinfo {eid} {064018}
  (\bibinfo {year} {2011})},\ \Eprint {http://arxiv.org/abs/1009.4876}
  {arXiv:1009.4876} \BibitemShut {NoStop}%
\bibitem [{\citenamefont {{Keidl}}\ \emph {et~al.}(2010)\citenamefont
  {{Keidl}}, \citenamefont {{Shah}}, \citenamefont {{Friedman}}, \citenamefont
  {{Kim}},\ and\ \citenamefont {{Price}}}]{Keidl-etal}%
  \BibitemOpen
  \bibfield  {author} {\bibinfo {author} {\bibfnamefont {T.~S.}\ \bibnamefont
  {{Keidl}}}, \bibinfo {author} {\bibfnamefont {A.~G.}\ \bibnamefont {{Shah}}},
  \bibinfo {author} {\bibfnamefont {J.~L.}\ \bibnamefont {{Friedman}}},
  \bibinfo {author} {\bibfnamefont {D.}~\bibnamefont {{Kim}}}, \ and\ \bibinfo
  {author} {\bibfnamefont {L.~R.}\ \bibnamefont {{Price}}},\ }\href {\doibase
  10.1103/PhysRevD.82.124012} {\bibfield  {journal} {\bibinfo  {journal}
  {\prd}\ }\textbf {\bibinfo {volume} {82}},\ \bibinfo {pages} {124012}
  (\bibinfo {year} {2010})},\ \Eprint {http://arxiv.org/abs/1004.2276}
  {arXiv:1004.2276} \BibitemShut {NoStop}%
\bibitem [{\citenamefont {{Barack}}\ and\ \citenamefont
  {{Ori}}(2000)}]{mode-sum-orig}%
  \BibitemOpen
  \bibfield  {author} {\bibinfo {author} {\bibfnamefont {L.}~\bibnamefont
  {{Barack}}}\ and\ \bibinfo {author} {\bibfnamefont {A.}~\bibnamefont
  {{Ori}}},\ }\href {\doibase 10.1103/PhysRevD.61.061502} {\bibfield  {journal}
  {\bibinfo  {journal} {\prd}\ }\textbf {\bibinfo {volume} {61}},\ \bibinfo
  {pages} {061502} (\bibinfo {year} {2000})},\ \Eprint
  {http://arxiv.org/abs/arXiv:gr-qc/9912010} {arXiv:gr-qc/9912010} \BibitemShut
  {NoStop}%
\bibitem [{\citenamefont {{Barack}}(2009)}]{Barack-review}%
  \BibitemOpen
  \bibfield  {author} {\bibinfo {author} {\bibfnamefont {L.}~\bibnamefont
  {{Barack}}},\ }\href {\doibase 10.1088/0264-9381/26/21/213001} {\bibfield
  {journal} {\bibinfo  {journal} {Class. and Quantum. Grav.}\ }\textbf
  {\bibinfo {volume} {26}},\ \bibinfo {pages} {213001} (\bibinfo {year}
  {2009})},\ \Eprint {http://arxiv.org/abs/arXiv:0908.1664} {arXiv:0908.1664}
  \BibitemShut {NoStop}%
\bibitem [{\citenamefont {{Barack}}\ \emph {et~al.}(2008)\citenamefont
  {{Barack}}, \citenamefont {{Ori}},\ and\ \citenamefont
  {{Sago}}}]{Barack-Ori-Sago}%
  \BibitemOpen
  \bibfield  {author} {\bibinfo {author} {\bibfnamefont {L.}~\bibnamefont
  {{Barack}}}, \bibinfo {author} {\bibfnamefont {A.}~\bibnamefont {{Ori}}}, \
  and\ \bibinfo {author} {\bibfnamefont {N.}~\bibnamefont {{Sago}}},\ }\href
  {\doibase 10.1103/PhysRevD.78.084021} {\bibfield  {journal} {\bibinfo
  {journal} {\prd}\ }\textbf {\bibinfo {volume} {78}},\ \bibinfo {pages}
  {084021} (\bibinfo {year} {2008})},\ \Eprint
  {http://arxiv.org/abs/arXiv:0808.2315} {arXiv:0808.2315} \BibitemShut
  {NoStop}%
\bibitem [{\citenamefont {{Hopper}}\ and\ \citenamefont
  {{Evans}}(2010)}]{Hopper-Evans}%
  \BibitemOpen
  \bibfield  {author} {\bibinfo {author} {\bibfnamefont {S.}~\bibnamefont
  {{Hopper}}}\ and\ \bibinfo {author} {\bibfnamefont {C.~R.}\ \bibnamefont
  {{Evans}}},\ }\href {\doibase 10.1103/PhysRevD.82.084010} {\bibfield
  {journal} {\bibinfo  {journal} {\prd}\ }\textbf {\bibinfo {volume} {82}},\
  \bibinfo {pages} {084010} (\bibinfo {year} {2010})},\ \Eprint
  {http://arxiv.org/abs/1006.4907} {arXiv:1006.4907} \BibitemShut {NoStop}%
\bibitem [{\citenamefont {{Barack}}\ and\ \citenamefont
  {{Sago}}(2009)}]{Barack-Sago-ISCO-shift}%
  \BibitemOpen
  \bibfield  {author} {\bibinfo {author} {\bibfnamefont {L.}~\bibnamefont
  {{Barack}}}\ and\ \bibinfo {author} {\bibfnamefont {N.}~\bibnamefont
  {{Sago}}},\ }\href {\doibase 10.1103/PhysRevLett.102.191101} {\bibfield
  {journal} {\bibinfo  {journal} {Phys. Rev. Lett.}\ }\textbf {\bibinfo
  {volume} {102}},\ \bibinfo {pages} {191101} (\bibinfo {year} {2009})},\
  \Eprint {http://arxiv.org/abs/arXiv:0902.0573} {arXiv:0902.0573} \BibitemShut
  {NoStop}%
\bibitem [{\citenamefont {{Barack}}\ \emph {et~al.}(2010)\citenamefont
  {{Barack}}, \citenamefont {{Damour}},\ and\ \citenamefont
  {{Sago}}}]{Barack-Damour-Sago}%
  \BibitemOpen
  \bibfield  {author} {\bibinfo {author} {\bibfnamefont {L.}~\bibnamefont
  {{Barack}}}, \bibinfo {author} {\bibfnamefont {T.}~\bibnamefont {{Damour}}},
  \ and\ \bibinfo {author} {\bibfnamefont {N.}~\bibnamefont {{Sago}}},\ }\href
  {\doibase 10.1103/PhysRevD.82.084036} {\bibfield  {journal} {\bibinfo
  {journal} {\prd}\ }\textbf {\bibinfo {volume} {82}},\ \bibinfo {pages}
  {084036} (\bibinfo {year} {2010})},\ \Eprint {http://arxiv.org/abs/1008.0935}
  {arXiv:1008.0935} \BibitemShut {NoStop}%
\bibitem [{\citenamefont {{Barack}}\ and\ \citenamefont
  {{Sago}}(2011)}]{Barack-Sago-precession}%
  \BibitemOpen
  \bibfield  {author} {\bibinfo {author} {\bibfnamefont {L.}~\bibnamefont
  {{Barack}}}\ and\ \bibinfo {author} {\bibfnamefont {N.}~\bibnamefont
  {{Sago}}},\ }\href {\doibase 10.1103/PhysRevD.83.084023} {\bibfield
  {journal} {\bibinfo  {journal} {\prd}\ }\textbf {\bibinfo {volume} {83}},\
  \bibinfo {eid} {084023} (\bibinfo {year} {2011})},\ \Eprint
  {http://arxiv.org/abs/1101.3331} {arXiv:1101.3331} \BibitemShut {NoStop}%
\bibitem [{\citenamefont {{Diaz-Rivera}}\ \emph {et~al.}(2004)\citenamefont
  {{Diaz-Rivera}}, \citenamefont {{Messaritaki}}, \citenamefont {{Whiting}},\
  and\ \citenamefont {{Detweiler}}}]{Diaz-Rivera}%
  \BibitemOpen
  \bibfield  {author} {\bibinfo {author} {\bibfnamefont {L.}\ \bibnamefont
  {{Diaz-Rivera}}}, \bibinfo {author} {\bibfnamefont {E.}~\bibnamefont
  {{Messaritaki}}}, \bibinfo {author} {\bibfnamefont {B.~F.}\ \bibnamefont
  {{Whiting}}}, \ and\ \bibinfo {author} {\bibfnamefont {S.}~\bibnamefont
  {{Detweiler}}},\ }\href {\doibase 10.1103/PhysRevD.70.124018} {\bibfield
  {journal} {\bibinfo  {journal} {\prd}\ }\textbf {\bibinfo {volume} {70}},\
  \bibinfo {pages} {124018} (\bibinfo {year} {2004})},\ \Eprint
  {http://arxiv.org/abs/arXiv:gr-qc/0410011} {arXiv:gr-qc/0410011} \BibitemShut
  {NoStop}%
\bibitem [{\citenamefont {{Quinn}}(2000)}]{Quinn}%
  \BibitemOpen
  \bibfield  {author} {\bibinfo {author} {\bibfnamefont {T.~C.}\ \bibnamefont
  {{Quinn}}},\ }\href {\doibase 10.1103/PhysRevD.62.064029} {\bibfield
  {journal} {\bibinfo  {journal} {\prd}\ }\textbf {\bibinfo {volume} {62}},\
  \bibinfo {pages} {064029} (\bibinfo {year} {2000})},\ \Eprint
  {http://arxiv.org/abs/arXiv:gr-qc/0005030} {arXiv:gr-qc/0005030} \BibitemShut
  {NoStop}%
\bibitem [{\citenamefont {{Detweiler}}\ and\ \citenamefont
  {{Whiting}}(2003)}]{Detweiler-Whiting}%
  \BibitemOpen
  \bibfield  {author} {\bibinfo {author} {\bibfnamefont {S.}~\bibnamefont
  {{Detweiler}}}\ and\ \bibinfo {author} {\bibfnamefont {B.~F.}\ \bibnamefont
  {{Whiting}}},\ }\href {\doibase 10.1103/PhysRevD.67.024025} {\bibfield
  {journal} {\bibinfo  {journal} {\prd}\ }\textbf {\bibinfo {volume} {67}},\
  \bibinfo {pages} {024025} (\bibinfo {year} {2003})},\ \Eprint
  {http://arxiv.org/abs/arXiv:gr-qc/0202086} {arXiv:gr-qc/0202086} \BibitemShut
  {NoStop}%
\bibitem [{\citenamefont {{Glampedakis}}\ and\ \citenamefont
  {{Kennefick}}(2002)}]{Glampedakis-Kennefick}%
  \BibitemOpen
  \bibfield  {author} {\bibinfo {author} {\bibfnamefont {K.}~\bibnamefont
  {{Glampedakis}}}\ and\ \bibinfo {author} {\bibfnamefont {D.}~\bibnamefont
  {{Kennefick}}},\ }\href {\doibase 10.1103/PhysRevD.66.044002} {\bibfield
  {journal} {\bibinfo  {journal} {\prd}\ }\textbf {\bibinfo {volume} {66}},\
  \bibinfo {pages} {044002} (\bibinfo {year} {2002})},\ \Eprint
  {http://arxiv.org/abs/arXiv:gr-qc/0203086} {arXiv:gr-qc/0203086} \BibitemShut
  {NoStop}%
\bibitem [{\citenamefont {{Carter}}(1968)}]{Carter}%
  \BibitemOpen
  \bibfield  {author} {\bibinfo {author} {\bibfnamefont {B.}~\bibnamefont
  {{Carter}}},\ }\href {\doibase 10.1103/PhysRev.174.1559} {\bibfield
  {journal} {\bibinfo  {journal} {Physical Review}\ }\textbf {\bibinfo {volume}
  {174}},\ \bibinfo {pages} {1559} (\bibinfo {year} {1968})}\BibitemShut
  {NoStop}%
\bibitem [{\citenamefont {{Brill}}\ \emph {et~al.}(1972)\citenamefont
  {{Brill}}, \citenamefont {{Chrzanowski}}, \citenamefont {{Pereira}},
  \citenamefont {{Fackerell}},\ and\ \citenamefont {{Ipser}}}]{Brill}%
  \BibitemOpen
  \bibfield  {author} {\bibinfo {author} {\bibfnamefont {D.~R.}\ \bibnamefont
  {{Brill}}}, \bibinfo {author} {\bibfnamefont {P.~L.}\ \bibnamefont
  {{Chrzanowski}}}, \bibinfo {author} {\bibfnamefont {C.~M.}\ \bibnamefont
  {{Pereira}}}, \bibinfo {author} {\bibfnamefont {E.~D.}\ \bibnamefont
  {{Fackerell}}}, \ and\ \bibinfo {author} {\bibfnamefont {J.~R.}\ \bibnamefont
  {{Ipser}}},\ }\href {\doibase 10.1103/PhysRevD.5.1913} {\bibfield  {journal}
  {\bibinfo  {journal} {\prd}\ }\textbf {\bibinfo {volume} {5}},\ \bibinfo
  {pages} {1913} (\bibinfo {year} {1972})}\BibitemShut {NoStop}%
\bibitem [{\citenamefont {{Bardeen}}\ \emph {et~al.}(1972)\citenamefont
  {{Bardeen}}, \citenamefont {{Press}},\ and\ \citenamefont
  {{Teukolsky}}}]{Bardeen}%
  \BibitemOpen
  \bibfield  {author} {\bibinfo {author} {\bibfnamefont {J.~M.}\ \bibnamefont
  {{Bardeen}}}, \bibinfo {author} {\bibfnamefont {W.~H.}\ \bibnamefont
  {{Press}}}, \ and\ \bibinfo {author} {\bibfnamefont {S.~A.}\ \bibnamefont
  {{Teukolsky}}},\ }\href {\doibase 10.1086/151796} {\bibfield  {journal}
  {\bibinfo  {journal} {\apj}\ }\textbf {\bibinfo {volume} {178}},\ \bibinfo
  {pages} {347} (\bibinfo {year} {1972})}\BibitemShut {NoStop}%
\bibitem [{\citenamefont {{Hughes}}(2000)}]{Hughes}%
  \BibitemOpen
  \bibfield  {author} {\bibinfo {author} {\bibfnamefont {S.~A.}\ \bibnamefont
  {{Hughes}}},\ }\href {\doibase 10.1103/PhysRevD.61.084004} {\bibfield
  {journal} {\bibinfo  {journal} {\prd}\ }\textbf {\bibinfo {volume} {61}},\
  \bibinfo {pages} {084004} (\bibinfo {year} {2000})}\BibitemShut {NoStop}%
\bibitem [{\citenamefont {{Hinderer}}\ and\ \citenamefont
  {{Flanagan}}(2008)}]{Hinderer-Flanagan}%
  \BibitemOpen
  \bibfield  {author} {\bibinfo {author} {\bibfnamefont {T.}~\bibnamefont
  {{Hinderer}}}\ and\ \bibinfo {author} {\bibfnamefont {{\'E}.~{\'E}.}\
  \bibnamefont {{Flanagan}}},\ }\href {\doibase 10.1103/PhysRevD.78.064028}
  {\bibfield  {journal} {\bibinfo  {journal} {\prd}\ }\textbf {\bibinfo
  {volume} {78}},\ \bibinfo {pages} {064028} (\bibinfo {year} {2008})},\
  \Eprint {http://arxiv.org/abs/0805.3337} {arXiv:0805.3337} \BibitemShut
  {NoStop}%
\bibitem [{\citenamefont {{Mino}}(2003)}]{Mino}%
  \BibitemOpen
  \bibfield  {author} {\bibinfo {author} {\bibfnamefont {Y.}~\bibnamefont
  {{Mino}}},\ }\href {\doibase 10.1103/PhysRevD.67.084027} {\bibfield
  {journal} {\bibinfo  {journal} {\prd}\ }\textbf {\bibinfo {volume} {67}},\
  \bibinfo {pages} {084027} (\bibinfo {year} {2003})},\ \Eprint
  {http://arxiv.org/abs/arXiv:gr-qc/0302075} {arXiv:gr-qc/0302075} \BibitemShut
  {NoStop}%
\bibitem [{GSL()}]{GSL}%
  \BibitemOpen
  \href@noop {} {}\bibinfo {note} {GNU Scientific Library,
  \url{http://www.gnu.org/software/gsl/}}\BibitemShut {NoStop}%
\bibitem [{\citenamefont {Misner}(1972)}]{Misner-superradiance}%
  \BibitemOpen
  \bibfield  {author} {\bibinfo {author} {\bibfnamefont {C.~W.}\ \bibnamefont
  {Misner}},\ }\href {\doibase 10.1103/PhysRevLett.28.994} {\bibfield
  {journal} {\bibinfo  {journal} {Phys. Rev. Lett.}\ }\textbf {\bibinfo
  {volume} {28}},\ \bibinfo {pages} {994} (\bibinfo {year} {1972})}\BibitemShut
  {NoStop}%
\bibitem [{\citenamefont {{Canizares}}\ \emph {et~al.}(2010)\citenamefont
  {{Canizares}}, \citenamefont {{Sopuerta}},\ and\ \citenamefont
  {{Jaramillo}}}]{Canizares-Sopuerta:eccentric}%
  \BibitemOpen
  \bibfield  {author} {\bibinfo {author} {\bibfnamefont {P.}~\bibnamefont
  {{Canizares}}}, \bibinfo {author} {\bibfnamefont {C.~F.}\ \bibnamefont
  {{Sopuerta}}}, \ and\ \bibinfo {author} {\bibfnamefont {J.~L.}\ \bibnamefont
  {{Jaramillo}}},\ }\href {\doibase 10.1103/PhysRevD.82.044023} {\bibfield
  {journal} {\bibinfo  {journal} {\prd}\ }\textbf {\bibinfo {volume} {82}},\
  \bibinfo {pages} {044023} (\bibinfo {year} {2010})},\ \Eprint
  {http://arxiv.org/abs/1006.3201} {arXiv:1006.3201} \BibitemShut {NoStop}%
\bibitem [{\citenamefont {{Haas}}(2007)}]{Haas}%
  \BibitemOpen
  \bibfield  {author} {\bibinfo {author} {\bibfnamefont {R.}~\bibnamefont
  {{Haas}}},\ }\href {\doibase 10.1103/PhysRevD.75.124011} {\bibfield
  {journal} {\bibinfo  {journal} {\prd}\ }\textbf {\bibinfo {volume} {75}},\
  \bibinfo {pages} {124011} (\bibinfo {year} {2007})},\ \Eprint
  {http://arxiv.org/abs/arXiv:0704.0797} {arXiv:0704.0797} \BibitemShut
  {NoStop}%
\bibitem [{\citenamefont {Haas}()}]{Hass-priv-comm}%
  \BibitemOpen
  \bibfield  {author} {\bibinfo {author} {\bibfnamefont {R.}~\bibnamefont
  {Haas}},\ }\href@noop {} {}\bibinfo {note} {Private
  communication}\BibitemShut {NoStop}%
\bibitem [{\citenamefont {Detweiler}()}]{Detweiler:priv_comm}%
  \BibitemOpen
  \bibfield  {author} {\bibinfo {author} {\bibfnamefont {S.}~\bibnamefont
  {Detweiler}},\ }\href@noop {} {}\bibinfo {note} {Private
  communication}\BibitemShut {NoStop}%
\bibitem [{\citenamefont {{Haas}}\ and\ \citenamefont
  {{Poisson}}(2005)}]{Haas-Poisson-mass_change}%
  \BibitemOpen
  \bibfield  {author} {\bibinfo {author} {\bibfnamefont {R.}~\bibnamefont
  {{Haas}}}\ and\ \bibinfo {author} {\bibfnamefont {E.}~\bibnamefont
  {{Poisson}}},\ }\href {\doibase 10.1088/0264-9381/22/15/008} {\bibfield
  {journal} {\bibinfo  {journal} {Class. Quantum. Grav.}\ }\textbf {\bibinfo
  {volume} {22}},\ \bibinfo {pages} {739} (\bibinfo {year} {2005})},\ \Eprint
  {http://arxiv.org/abs/arXiv:gr-qc/0411108} {arXiv:gr-qc/0411108} \BibitemShut
  {NoStop}%
\bibitem [{\citenamefont {{Burko}}\ \emph {et~al.}(2002)\citenamefont
  {{Burko}}, \citenamefont {{Harte}},\ and\ \citenamefont
  {{Poisson}}}]{Burko-Harte-Poisson}%
  \BibitemOpen
  \bibfield  {author} {\bibinfo {author} {\bibfnamefont {L.~M.}\ \bibnamefont
  {{Burko}}}, \bibinfo {author} {\bibfnamefont {A.~I.}\ \bibnamefont
  {{Harte}}}, \ and\ \bibinfo {author} {\bibfnamefont {E.}~\bibnamefont
  {{Poisson}}},\ }\href {\doibase 10.1103/PhysRevD.65.124006} {\bibfield
  {journal} {\bibinfo  {journal} {\prd}\ }\textbf {\bibinfo {volume} {65}},\
  \bibinfo {pages} {124006} (\bibinfo {year} {2002})},\ \Eprint
  {http://arxiv.org/abs/arXiv:gr-qc/0201020} {arXiv:gr-qc/0201020} \BibitemShut
  {NoStop}%
\end{thebibliography}

%

\end{document}